\documentclass[twocolumn]{aastex62} 

\usepackage{hyperref,enumerate}
\input{colordvi}

\hypersetup{
    unicode=false,                  
    pdftoolbar=true,                
    pdfmenubar=true,                
    pdffitwindow=true,              
    pdfstartview={FitH},            
    pdftitle={R443130},             
    pdfauthor={vplacco},            
    pdfsubject={Astronomy},         
    pdfcreator={dvipdf},            
    pdfproducer={dvipdf},           
    pdfkeywords={metal-poor stars}, 
    pdfnewwindow=true,              
    colorlinks=true,                
    linkcolor=red,                  
    citecolor=blue,                 
    filecolor=magenta,              
    urlcolor=cyan,                  
    breaklinks=true,
    linktocpage
}

\newcommand{\abund}[2]{\ensuremath{[\mathrm{#1}/\mathrm{#2}]}}
\newcommand{\afe}{\abund{\alpha}{Fe}}
\newcommand{\cfe}{\abund{C}{Fe}}

\newcommand{\xfe}[1]{\abund{#1}{Fe}}
\newcommand{\metal}{\abund{Fe}{H}}

\newcommand{\teff}{\ensuremath{T_\mathrm{eff}}}
\newcommand{\logg}{\ensuremath{\log\,g}}

\received{March 15, 2018}
\revised{April 27, 2018}
\accepted{April 30, 2018}

\shorttitle{Metal-Poor Stars from RAVE}
\shortauthors{Placco et al.}

\begin{document}

\title{Spectroscopic Validation of Low-Metallicity Stars from RAVE}

\author[0000-0003-4479-1265]{Vinicius M.\ Placco}
\altaffiliation{Visiting astronomer, Kitt Peak National Observatory.}
\affiliation{Department of Physics, University of Notre Dame, Notre Dame, IN 46556, USA}
\affiliation{JINA Center for the Evolution of the Elements, USA}

\author[0000-0003-4573-6233]{Timothy C.\ Beers}
\affiliation{Department of Physics, University of Notre Dame, Notre Dame, IN 46556, USA}
\affiliation{JINA Center for the Evolution of the Elements, USA}

\author[0000-0002-7529-1442]{Rafael M.\ Santucci}
\affiliation{Instituto de Estudos S\'ocio-Ambientais, Planet\'ario, 
Universidade Federal de Goi\'as, Goi\^ania, GO 74055-140, Brazil}
\affiliation{Instituto de F\'isica, Universidade Federal de Goi\'as, Campus
Samambaia, Goi\^ania, GO 74001-970, Brazil}

\author{Julio Chanam\'e}
\affiliation{Instituto de Astrof\'isica, Pontificia Universidad Cat\'olica de
Chile, Santiago, Chile}
\affiliation{Millenium Institute of Astrophysics, Santiago, Chile}

\author{Mar\'ia Paz Sep\'ulveda}
\affiliation{Instituto de Astrof\'isica, Pontificia Universidad Cat\'olica de
Chile, Santiago, Chile}
\affiliation{Millenium Institute of Astrophysics, Santiago, Chile}

\author{Johanna Coronado}
\affiliation{Max-Planck-Institut f\"ur Astronomie, 
K\"oningstuhl 17, D-69117 Heidelberg, Germany}
\affiliation{Instituto de Astrof\'isica, Pontificia Universidad Cat\'olica de
Chile, Santiago, Chile}

\author{Sean D.\ Points}
\affiliation{Cerro Tololo Inter-American Observatory, Casilla 603, La Serena, Chile}

\author{Catherine C.\ Kaleida}
\affiliation{Space Telescope Science Institute, Baltimore, MD 21218, USA}

\author[0000-0001-7479-5756]{Silvia Rossi}
\affiliation{Instituto de Astronomia,  Geof\'{i}sica e Ci\^{e}ncias Atmosf\'{e}ricas,
Universidade de S\~{a}o Paulo, SP 05508-900, Brazil}

\author{Georges Kordopatis}
\affiliation{Laboratoire Lagrange, Universit\'e C\^ote d'Azur, Observatoire de
la C\^ote d'Azur, CNRS, F-06304 Nice cedex 4, France}

\author{Young Sun Lee}
\affiliation{Department of Astronomy and Space Science, Chungnam National
University, Daejeon 34134, Korea}

\author{Gal Matijevi\v{c}}
\affiliation{Leibniz Institut f\"{u}r Astrophysik Potsdam (AIP), An der
Sterwarte 16, D-14482 Potsdam, Germany}

\author[0000-0002-2139-7145]{Anna Frebel}
\affiliation{Department of Physics and Kavli Institute for Astrophysics and
Space Research, \\ Massachusetts Institute of Technology, Cambridge, MA 02139, USA}
\affiliation{JINA Center for the Evolution of the Elements, USA}

\author[0000-0001-6154-8983]{Terese T.\ Hansen}
\affiliation{Observatories of the Carnegie Institution of Washington, 
Pasadena, CA 91101, USA}

\author[0000-0002-5463-6800]{Erika M.\ Holmbeck}
\affiliation{Department of Physics, University of Notre Dame, Notre Dame, IN 46556, USA}
\affiliation{JINA Center for the Evolution of the Elements, USA}

\author[0000-0002-0470-0800]{Kaitlin C.\ Rasmussen}
\affiliation{Department of Physics, University of Notre Dame, Notre Dame, IN 46556, USA}
\affiliation{JINA Center for the Evolution of the Elements, USA}

\author[0000-0001-5107-8930]{Ian U.\ Roederer}
\affiliation{Department of Astronomy, University of Michigan, Ann Arbor, MI 48109, USA}
\affiliation{JINA Center for the Evolution of the Elements, USA}

\author[0000-0002-5095-4000]{Charli M.\ Sakari}
\affiliation{Department of Astronomy, University of Washington, Seattle, WA 98195-1580, USA}

\author[0000-0002-9594-6143]{Devin D.\ Whitten}
\affiliation{Department of Physics, University of Notre Dame, Notre Dame, IN 46556, USA}
\affiliation{JINA Center for the Evolution of the Elements, USA}

\correspondingauthor{Vinicius M.\ Placco}
\email{vplacco@nd.edu}

\begin{abstract}

We present results from a medium-resolution ($R\sim2,000$)
spectroscopic follow-up campaign of 1,694 bright ($V<13.5$), very metal-poor
star candidates from the RAdial Velocity Experiment (RAVE).  Initial selection
of the low-metallicity targets was based on the stellar parameters published in RAVE Data
Releases 4 and 5. Follow-up was accomplished with the Gemini-N and Gemini-S, the
ESO/NTT, the KPNO/Mayall, and the SOAR telescopes. The wavelength coverage for
most of the observed spectra allows for the determination of carbon and
$\alpha$-element abundances, which are crucial for considering the nature and
frequency of the carbon-enhanced metal-poor (CEMP) stars in this sample.  We
find that 88\% of the observed stars have \metal\,$\leq -1.0$, 61\% have
\metal\,$\leq -2.0$, and 3\% have \metal\,$\leq -3.0$ (with four stars at
\metal\,$\leq -3.5$).  There are 306 CEMP star candidates in this
sample, and  we identify 169 CEMP Group~I, 131 CEMP Group~II, and 6 CEMP
Group~III stars from the $A$(C) vs. [Fe/H] diagram.  Inspection of the
\abund{\alpha}{C} abundance ratios reveals that five of the CEMP Group~II stars
can be classified as ``mono-enriched second-generation'' stars.  Gaia DR1
matches were found for 734 stars, and we show that transverse velocities can be
used as a confirmatory selection criteria for low-metallicity candidates.  
Selected stars from our validated list are being followed-up with
high-resolution spectroscopy, to reveal their full chemical abundance
patterns for further studies.

\end{abstract}


\section{Introduction}
\label{intro}

Low-metallicity stars provide a direct window on the origin of the first
(Population III) stars to form in the Universe, and on the chemical and dynamical
evolution of the Milky Way \citep{beers2005,frebel2015,chiaki2017,jeon2017}.
In particular, as sample sizes have grown over the last few decades, interest
has focused on the nature of the very metal-poor (VMP;
[Fe/H]\footnote{\abund{A}{B} = $log(N_A/{}N_B)_{\star} - log(N_A/{}N_B)
_{\odot}$, where $N$ is the number density of atoms of a given element in the
star ($\star$) and the Sun ($\odot$), respectively.}\,$< -2.0$), extremely
metal-poor (EMP; [Fe/H]\,$< -3.0$), and ultra metal-poor (UMP; [Fe/H]\,$< -4.0$)
stars.
Detailed spectroscopic studies of these objects are the best way to identify
and distinguish between a number of possible scenarios for the enrichment of
early star-forming gas clouds soon after the Big Bang \citep[see,
e.g.,][]{norris2013,yong2013,hansen2014,placco2014,roederer2014,frebel2015b,hansen2015,placco2015,placco2016b}.

Furthermore, it has been recognized that carbon is ubiquitous in the early
universe, based on empirical evidence that the frequencies of carbon-enhanced
metal-poor \citep[CEMP; \cfe\,$ \geq\,+0.7$, e.g., ][]{beers2005,aoki2007}
stars increase
with decreasing stellar metallicity, from 20\% for VMP stars to at
least 80\% for UMP stars \citep{placco2014c}. 
The full elemental-abundance patterns for CEMP stars are required in order to
probe the nature of the different progenitor populations responsible for the
production of carbon and other elements.  Recent studies \citep[e.g.,][and
references therein]{yoon2016} show that the majority of CEMP stars with
\metal\,$ \leq -3.0$ belong to the CEMP-no sub-class, characterized by a lack of
enhancements in the neutron-capture elements (e.g., [Ba/Fe] $<$ 0.0).  The
brightest EMP star in the sky, BD$+$44:493, with \metal\,$= -3.8$ and $V=9.0$,
is a CEMP-no star \citep{ito2013,placco2014b}, and shares a common light-element
abundance signature with the most iron-poor star observed to date
\cite[\metal\,$\leq\,-7.5$;][]{keller2014,bessell2015}. The distinctive CEMP-no
pattern has also been identified in high-$z$ damped Lyman-$\alpha$ systems
\citep{cooke2012,cooke2014}, and is common among stars in the ultra-faint dwarf
spheroidal galaxies such as Segue-1 \citep{frebel2014}.  These, and other
observations, suggest that CEMP-no stars exhibit the nucleosynthesis products of
the first generation of stars \citep{sharma2017,hartwig2018}.

Another important subclass of metal-poor stars are the objects that exhibit
over-abundances in elements synthesized by the rapid neutron-capture process
\citep[$r$-process;][]{b2fh,cameron1957}. The so-called $r$-II stars are
identified by enhancements in europium (\xfe{Eu}\,$\geq +1.0$) and a low Ba
abundance relative to Eu \citep[\abund{Ba}{Eu}\,$\leq 0.0$;][]{beers2005}.
These stars are believed to be formed preferentially in ultra-faint dwarf
galaxies \citep{ji2016,roederer2016b}, and their abundance patterns have been
suggested to arise from the nucleosynthesis products of a neutron star merger
\citep{lattimer1974}. Recent photometric and spectroscopic observations of the
electromagnetic counterpart of the gravitational wave event GW170817 has
confirmed this association \citep{abbott2017,drout2017,shappee2017}.

The first $r$-II star, CS-22892-052, was identified by \citet{sneden1994}, based
on high-resolution spectroscopic follow-up of VMP/EMP stars discovered in the HK
survey \citep{beers1985,beers1992}. Over time, the numbers of known $r$-II stars
slowly grew.  Dedicated observational efforts, beginning with the work of
\citet{christlieb2004} and \citet{barklem2005}, and more recently, by
\citet{roederer2014d}, and others (typically one or two stars at a time), have
been able to identify a total of $\sim 25-30$ $r$-II stars, which account for
roughly 3\% of the population of VMP stars.  Most recently, the $R$-Process
Alliance (RPA) was established \citep[e.g.][]{hansen2018}, with the aim
to at least quadruple the number of known $r$-II stars over the next few years,
and to study their abundance patterns at high spectral resolution.

Over the last 25 years, large-scale survey efforts have dramatically increased the
numbers of known low-metallicity stars in the Galaxy, enabling their further study
with high-resolution spectroscopy.  Together, the HK survey
\citep{beers1985,beers1992} and the Hamburg/ESO Survey
\citep[HES;][]{frebel2006,christlieb2008} identified several thousand VMP stars. To date,
two of the three most metal-poor stars found in the halo of the Galaxy,
HE~0107$-$5240 \citep[\metal = $-$5.2;][]{christlieb2002} and HE~1327$-$2326
\citep[\metal = $-$5.6;][]{frebel2005}, were first identified from spectroscopic
follow-up of candidates in the HES database. These numbers have been further
increased, through the use of medium-resolution spectroscopy carried out during
the SDSS \citep{york2000} and the sub-surveys Sloan Extension for Galactic
Understanding and Exploration \citep[SEGUE-1;][]{yanny2009} and SEGUE-2, to many
tens of thousands of VMP (and $\sim$1,000 EMP) stars.

However, due to limitations in the input source catalogs (e.g., saturation of
the HK/HES prism plates, and the $g \sim 14$ bright limit from SDSS scans), the
above surveys have not been able to provide large numbers of VMP/EMP stars with
$V < 12$. Yet, bright examples of such stars are the {\it best available
targets} for high-resolution spectroscopic follow-up, since they can be readily
observed to very high S/N ratios with 4m- and 8m-class ground-based telescopes,
and are the {\it only stars} that can be observed at high spectral resolution
in the near-UV with the Hubble Space Telescope
\citep[e.g.,][]{sneden1998,roederer2012c,roederer2012d,
placco2014b,roederer2014c,placco2015b,roederer2016}.

The RAVE \citep[RAdial Velocity Experiment;][]{steinmetz2006} survey has broken
through this limitation, and has determined atmospheric parameters and accurate
radial velocities for a magnitude-limited ($ 9 < I < 12$) sample of over
400,000 stars in the Southern Hemisphere \citep[DR4;][]{kordopatis2013}, based
on moderate-resolution spectroscopy ($R \sim 8,000$). Even though RAVE provides
reasonably reliable atmospheric-parameter estimates (T$_{\rm eff}$, log $g$,
and \metal), it is not possible for RAVE to measure \cfe, as prominent
carbon-related features lie outside their narrow spectral region, centered on
the \ion{Ca}{1} triplet at $\sim 8,500$\,{\AA}\footnote{Other ongoing surveys,
including the Best \& Brightest survey of \citet{schlaufman2014} and the
SkyMapper survey \citep{keller2007}, have also identified large numbers of
bright VMP candidate stars. Medium- and high-resolution spectroscopy of stars
from these surveys have been collected as part of the RPA effort, and
will be reported on in due course.}.

\begin{figure*}[!ht]
\epsscale{1.05}
\plotone{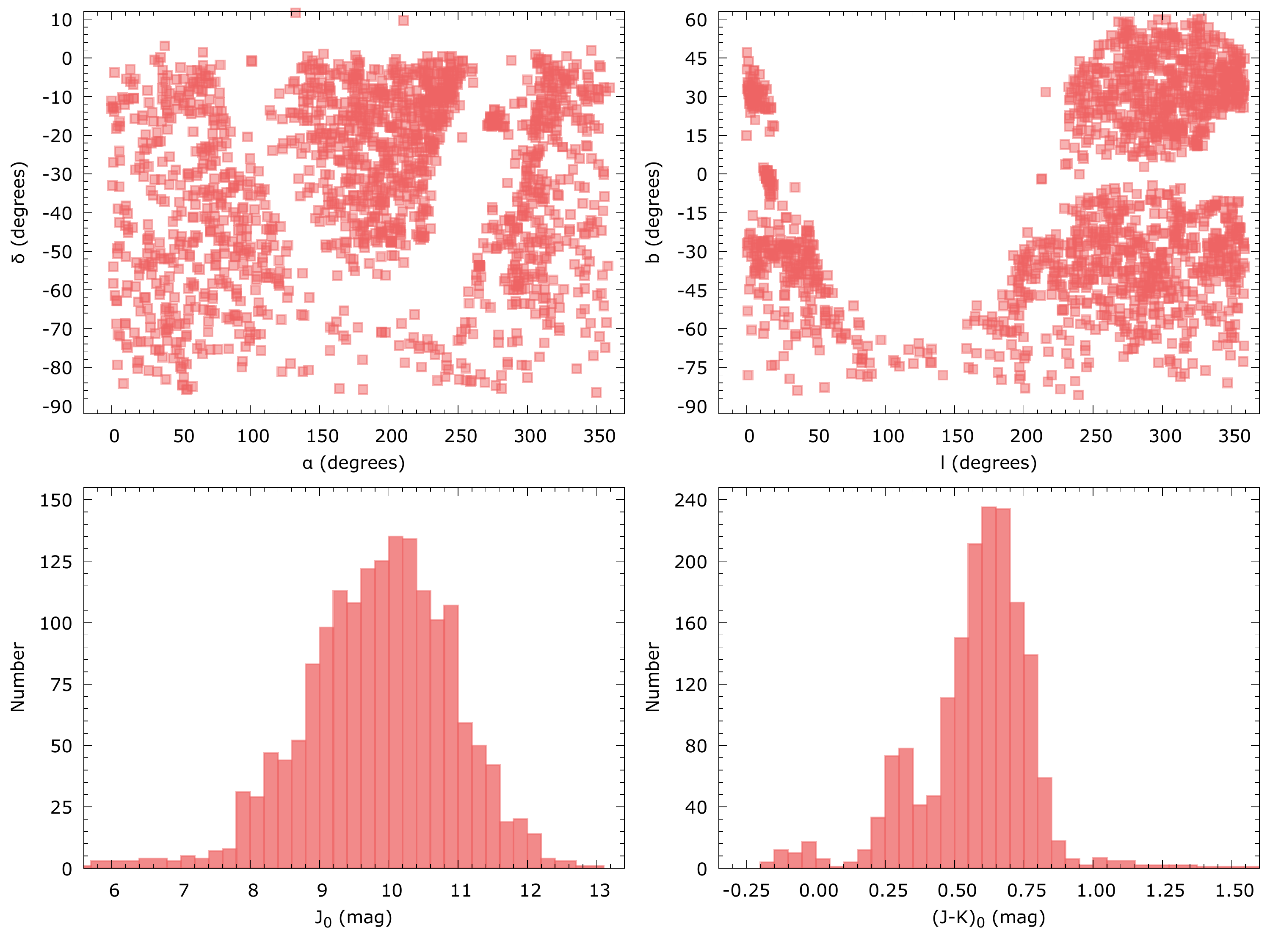}
\caption{Upper panels: Equatorial and Galactic coordinates for the observed
targets. Lower panels: Distributions of absorption-corrected $J_0$ magnitudes
and de-reddened $(J-K)_0$ colors.}
\label{coords}
\end{figure*}

In this paper, we report on the medium-resolution ($R \sim 2,000$) spectroscopic
follow-up of VMP star candidates selected from RAVE.  The main goals are to
determine carbon abundances for a large sample of metal-poor stars from RAVE,
and to validate the published RAVE atmospheric-parameter estimates. Once
identified, interesting targets are re-observed with high-resolution
spectroscopy, in order to determine their full chemical-abundance patterns. 
This paper is outlined as follows. Section~\ref{secobs} describes the target
selection for the medium-resolution spectroscopic investigation and the
follow-up observations, followed by the determinations of the stellar
atmospheric parameters and abundances in Section~\ref{secatm}. We provide a
comparison between the RAVE parameters and our determinations in
Section~\ref{secrave}, and present Gaia DR1-based distances and proper motions
for a subset of our targets in Section~\ref{gaias}.  Section~\ref{seccomp}
describes the importance of the carbon and $\alpha$-element abundances in
further selecting targets for high-resolution follow-up. Our concluding remarks
are provided in Section~\ref{final}.

\section{Target Selection and Observations}
\label{secobs}

The final data release of the RAVE \citep[DR5;][]{kunder2017} presents
atmospheric parameters, radial velocities, individual abundances, and distances
for 520,701 stars, mostly in the $9 < I < 12$ magnitude range. Due to its
limited spectral range (8410-8795\,{\AA}), it is not possible to determine
carbon abundances for these stars based on RAVE spectra alone.  There are
several advantages in working with RAVE DR5 data: (i) Their relatively
reliable [Fe/H] estimates help in selecting only metal-poor candidates,
and avoid color-dependent effects that can compromise this kind of
selection, and (ii) The radial velocities of the RAVE stars have an average
error of no more than a few km s$^{-1}$. These velocities, together with
reliable proper motions and parallaxes from the Gaia mission, will allow for
precision determinations of the full space motions that are required for further
kinematic analysis.

\begin{figure*}[!ht]
\epsscale{1.10}
\plotone{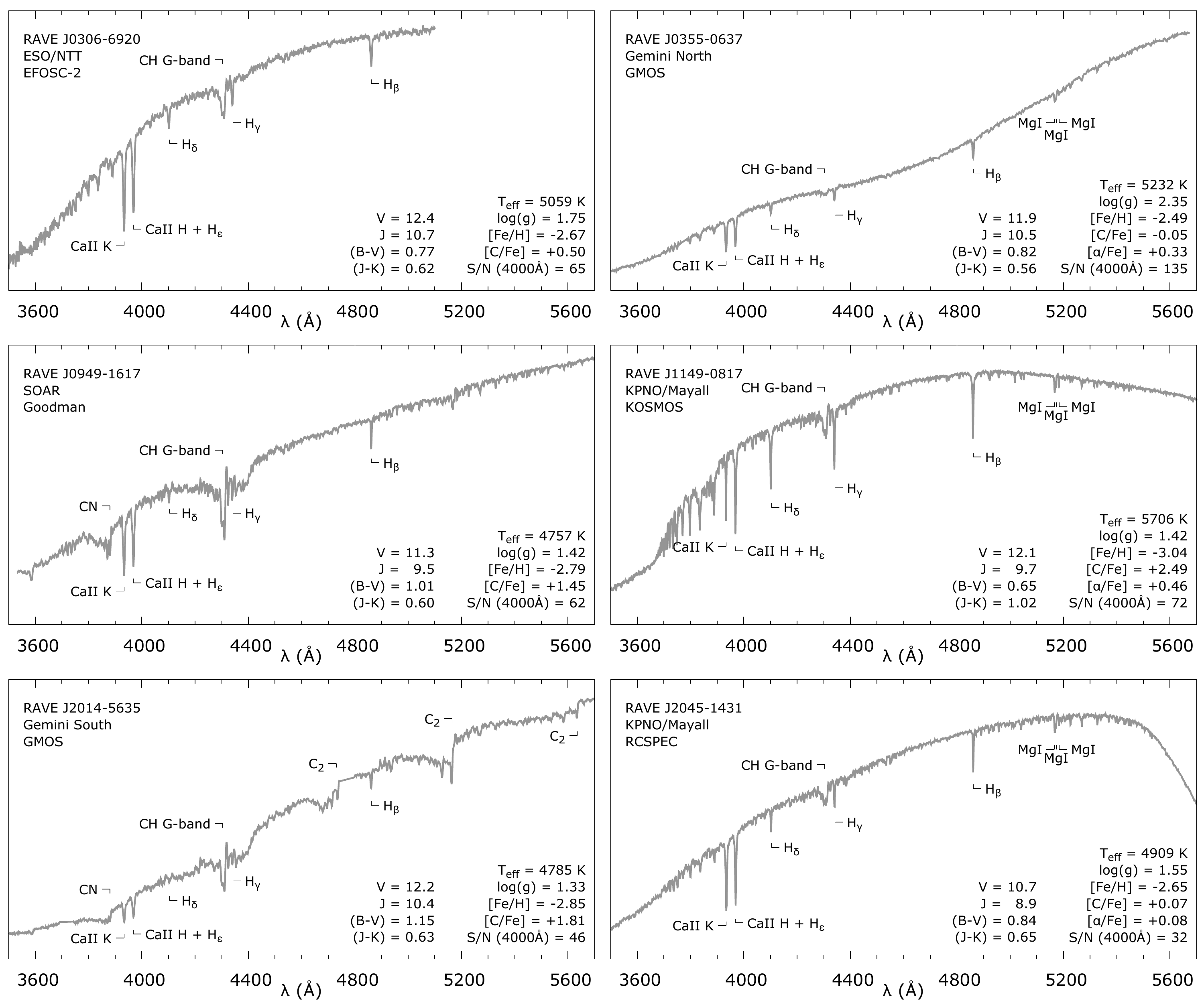}
\caption{Example medium-resolution spectra for six RAVE stars observed with the
resources described in Section~\ref{subobs}. The magnitudes, colors, and estimated 
parameters are listed in the lower-right part of each panel (see details in
Section~\ref{secatm}). Prominent absorption features are identified in each
spectrum.}
\label{medspec}
\end{figure*}

\subsection{Target Selection from the RAVE Database}

RAVE presents the ideal sample for our proposed search for bright
low-metallicity stars with medium-resolution spectroscopy. 
Since RAVE provides stellar parameters,  we could be quite selective on the
temperatures and metallicities of each target before carrying out our 
spectroscopic follow-up, roughly covering the wavelength range
[3500:5500]\,{\AA}.

There are three primary goals: (i) Obtain \cfe\, and \afe\, estimates for a
large number of metal-poor stars in RAVE, (ii) Derive estimates of the
fractions of CEMP stars, as a function of \metal, for a large unbiased sample
of halo stars, selected without a-priori knowledge of the likelihood of carbon
enhancement, and (iii) Determine, on the basis of the full space motions of
these stars (when available), the fractions of CEMP stars associated
with the inner- and outer-halo populations of the Galaxy.


The low-metallicity candidates were selected from the RAVE DR4 and DR5 catalogs.
The selection criteria were applied to the RAVE stellar parameters ($3,500 <$
\teff (K)\footnote{{\texttt{Teff\_K}} for DR4 and {\texttt{Teff\_N\_K}} for DR5}
$< 7,000$; \metal\footnote{{\texttt{c[M/H]\_K}} for DR4 and {\texttt{Met\_N\_K}}
for DR5}\,$ \leq -1.5$), and 2MASS photometric quality flags
\citep[{\texttt{ph\_qual}}$_{\rm JHK}$ = AAA;][]{skrutskie2006}. We also
tracked (but did not use for the selection) the internal quality flag from
the RAVE stellar parameter pipeline ({\texttt{QK - Algo\_Conv\_K}}), the
signal-to-noise ratio for two different pipeline implementations
({\texttt{STN\_SPARV}} and {\texttt{SNR\_K}}), and the first three
morphological flags \citep[{\texttt{c1/c2/c3}};][]{matijevic2012} for each
star. These were used to assess possible discrepancies between the RAVE
parameters and the values determined by our spectroscopic follow-up (see
Section~\ref{secrave} for further details).  That is, we sought to explore
which of the RAVE flags should be used to exclude potential contaminating
stars in our search, and which could be ignored.

\subsection{Medium-resolution Follow-up Observations}
\label{subobs}

The medium-resolution spectroscopic follow-up campaign was conducted from semesters 
2014A to 2017A, and collected 1,835 spectra of 1,694 unique metal-poor candidates. We
used six different telescope/instrument setups: 
(i)   SOAR/Goodman, 
(ii)  Gemini North/GMOS-N,
(iii) Gemini South/GMOS-S,
(iv)  Mayall/RCSPEC,
(v)   Mayall/KOSMOS, and
(vi)  NTT/EFOSC-2.
Table~\ref{telescope} lists the object name, observation date, telescope, instrument,
program ID, and exposure time for the observed candidates. 
Table~\ref{coordsmags} lists their coordinates, magnitudes, color indices, and
reddening estimates \citep{schlegel1998}.
Figure~\ref{coords} shows the equatorial and Galactic coordinates of the
observed targets, as well as the distribution of their extinction-corrected
$J_0$ magnitudes and de-reddened $(J-K)_0$ color indices.

Details on each observing setup are provided below. For consistency across the
different instruments, we chose grating/slit combinations that would yield a
resolving power $R\sim 1,200-2,000$, and exposure times sufficient to
reach a signal-to-noise ratio of at least S/N\,$ \sim 40$ per pixel at the
\ion{Ca}{2}~K line (3933.3\,{\AA}). 
The average value for the 1,694 observed spectra is S/N\,$ \sim 54$ at 4000\,{\AA}.
Calibration frames included arc-lamp
exposures, bias frames, and quartz flats. All tasks related to spectral
reduction, extraction, and wavelength calibration were performed using standard
IRAF\footnote{\href{http://iraf.noao.edu}{http://iraf.noao.edu}.} packages
\citep[see][for further details]{placco2013}.

\begin{figure*}[!ht]
\epsscale{1.075}
\plotone{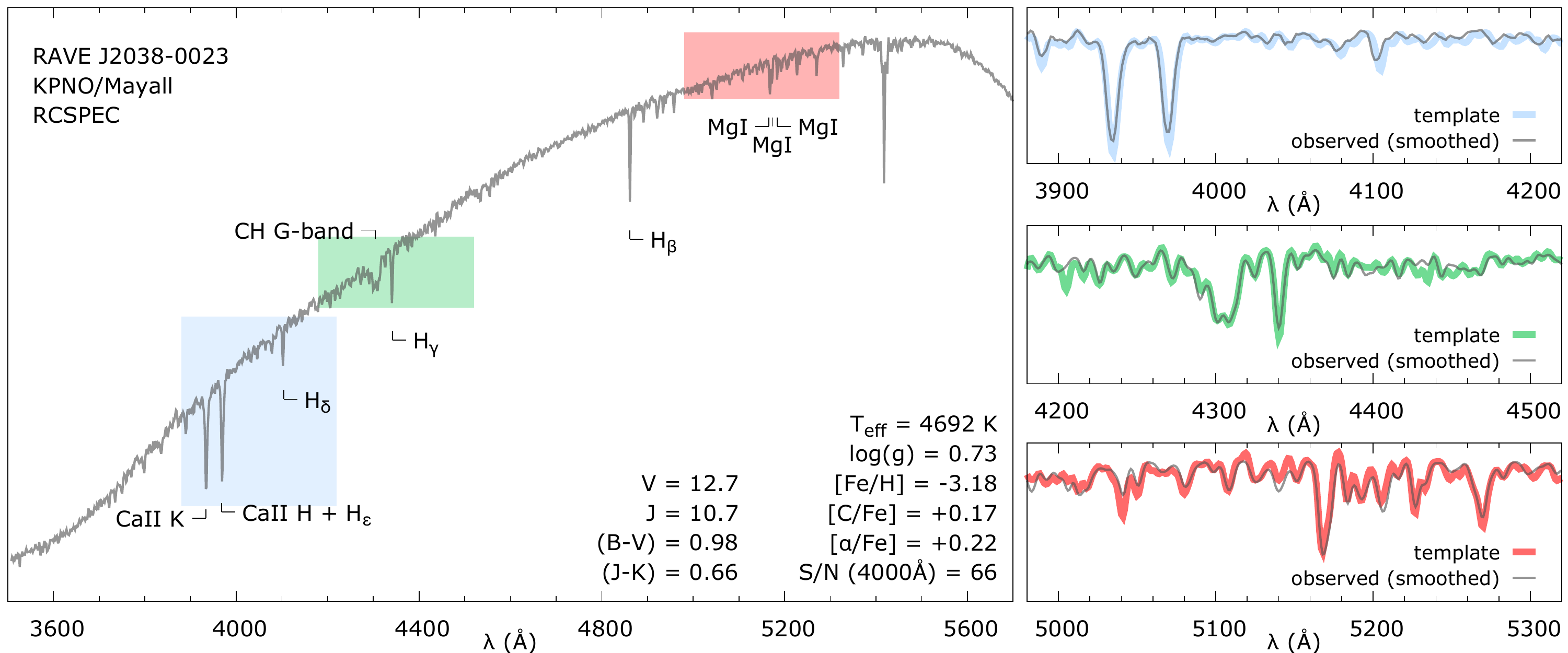}
\caption{Left panel: Example RCSPEC spectrum in the wavelength range as analyzed by
the n-SSPP, with key absorption features identified.
Right panels: Comparison between the observed and the synthetic spectra,
generated using the parameters listed on the left panel.  The \ion{Ca}{2}~K
line (top right) is used to determine \metal; \cfe\ is determined from the
CH $G$-band (middle right); and the \ion{Mg}{1} triplet (lower right) is used to
estimate \afe.} 
\label{nsspp} 
\end{figure*}

\paragraph{SOAR Telescope}

335 stars were observed with the 4.1\,m Southern Astrophysical
Research (SOAR) telescope. 
The Goodman Spectrograph was used with the
600~l~mm$^{\rm{-1}}$ grating, the blue setting, and a 1$\farcs$03 slit,
resulting in a wavelength coverage in the range [3600:6200]\,{\AA} at resolving power $R \sim 1,500$. 
An example Goodman spectrum can be seen on the middle-left panel of
Figure~\ref{medspec}.

\paragraph{Gemini North and South Telescopes}

192 stars were observed with the twin 8.1\,m Gemini North (30
stars) and Gemini South (162 stars) telescopes. In both cases, we used the
B600~l~mm$^{\rm{-1}}$ grating (G5323 for GMOS South and G5307 for GMOS North)
and a 1$\farcs$00 slit, 
resulting in a wavelength coverage in the range [3200:5800]\,{\AA} at resolving power $R \sim 2,000$. 
Example GMOS-N and GMOS-S spectra can be seen in the top-right and bottom-left
panels of Figure~\ref{medspec}, respectively.

\paragraph{KPNO Mayall Telescope}

444 stars were observed with the 4\,m Mayall telescope, located at
Kitt Peak National Observatory, using the R-C spectrograph (337 stars) and
the KOSMOS spectrograph (107 stars).
For the R-C spectrograph, we used the KPC007 grating (632~l~mm$^{\rm{-1}}$),
the blue setting, and a 1$\farcs$0 slit, 
resulting in a wavelength coverage in the range [3500:6000]\,{\AA} at resolving power $R \sim 1,500$. 
For the KOSMOS spectrograph, we used the 600~l~mm$^{\rm{-1}}$ grating, the blue
setting, and a 0$\farcs$9 slit, 
resulting in a wavelength coverage in the range [3600:6300]\,{\AA} at resolving power $R \sim 1,800$. 
Example RCSPEC and KOSMOS spectra can be seen in the bottom-right and
middle-right panels of Figure~\ref{medspec}, respectively.

\paragraph{ESO New Technology Telescope}

723 stars were observed with the 3.58\,m New Technology Telescope
(NTT), located at La Silla Observatory, part of the European Southern
Observatory.
We used the EFOSC-2 spectrograph with Grism\#7 (600~gr~mm$^{\rm{-1}}$) and a 1$\farcs$0 slit, 
resulting in a wavelength coverage in the range [3300:5100]\,{\AA} at resolving power $R \sim 1,200$. 
An example EFOSC-2 spectrum can be seen in the top-left panel of
Figure~\ref{medspec}.

\section{Stellar Parameters and Abundances}
\label{secatm}

The stellar atmospheric parameters, as well as carbon and $\alpha$-element
abundances, were determined using the n-SSPP \citep{beers2014,beers2017}, a
modified version of the SEGUE Stellar Parameter Pipeline
\citep[SSPP;][]{lee2008a,lee2008b,lee2013}.
Figure~\ref{nsspp} shows an example processing of the n-SSPP for the spectrum
of RAVE~J2038$-$0023\footnote{The parameters determined from the n-SSPP for
this star motivated us to obtain high-resolution spectroscopic follow-up using
MIKE/Magellan.  RAVE~J2038$-$0023 was then identified as the first
$r$-process-enhanced star identified in the RAVE database \citep{placco2017}.},
observed with RCSPEC on KPNO/Mayall.
The left panel shows the entire wavelength range, and identifies key absorption
features used for parameter estimation.
The right panels show comparisons between the observed spectrum and the
synthetic spectrum generated by the n-SSPP using the parameters quoted on the
left panel (i.e., the synthetic spectrum is not a fit, but a prediction).  The
\ion{Ca}{2}~K line (top right) is used to determine \metal; \cfe\ is determined
from the CH $G$-band (middle right); and the \ion{Mg}{1} triplet (lower right)
is used to estimate \afe. 

Atmospheric parameters were determined for $\sim 95\%$ of the observed sample
(1,614 out of 1,694 stars). The non-determinations arise from lack of
temperature estimates by the n-SSPP, due mostly to stars outside the [4000:7000]
\teff\ range, or stars with core emission in the \ion{Ca}{2}~K line.  The \cfe\
and \afe\ abundance ratios were calculated for 1,606 and \replaced{945}{849} stars,
respectively.  The carbon-abundance determination is not carried out for spectra
with S/N$<$10, and/or when high temperature renders the CH $G$-band molecular
feature too weak to be useful \citep[e.g.,][]{placco2016}. In addition, \afe\
could not be calculated for the NTT/EFOSC-2 spectra, due to the lack of spectral
coverage. 
The final parameters and abundances for the sample are listed in
Table~\ref{comprave}, as well as radial velocities and atmospheric parameters
from RAVE DR4 and DR5. 

\begin{figure}[!ht]
\epsscale{1.10}
\plotone{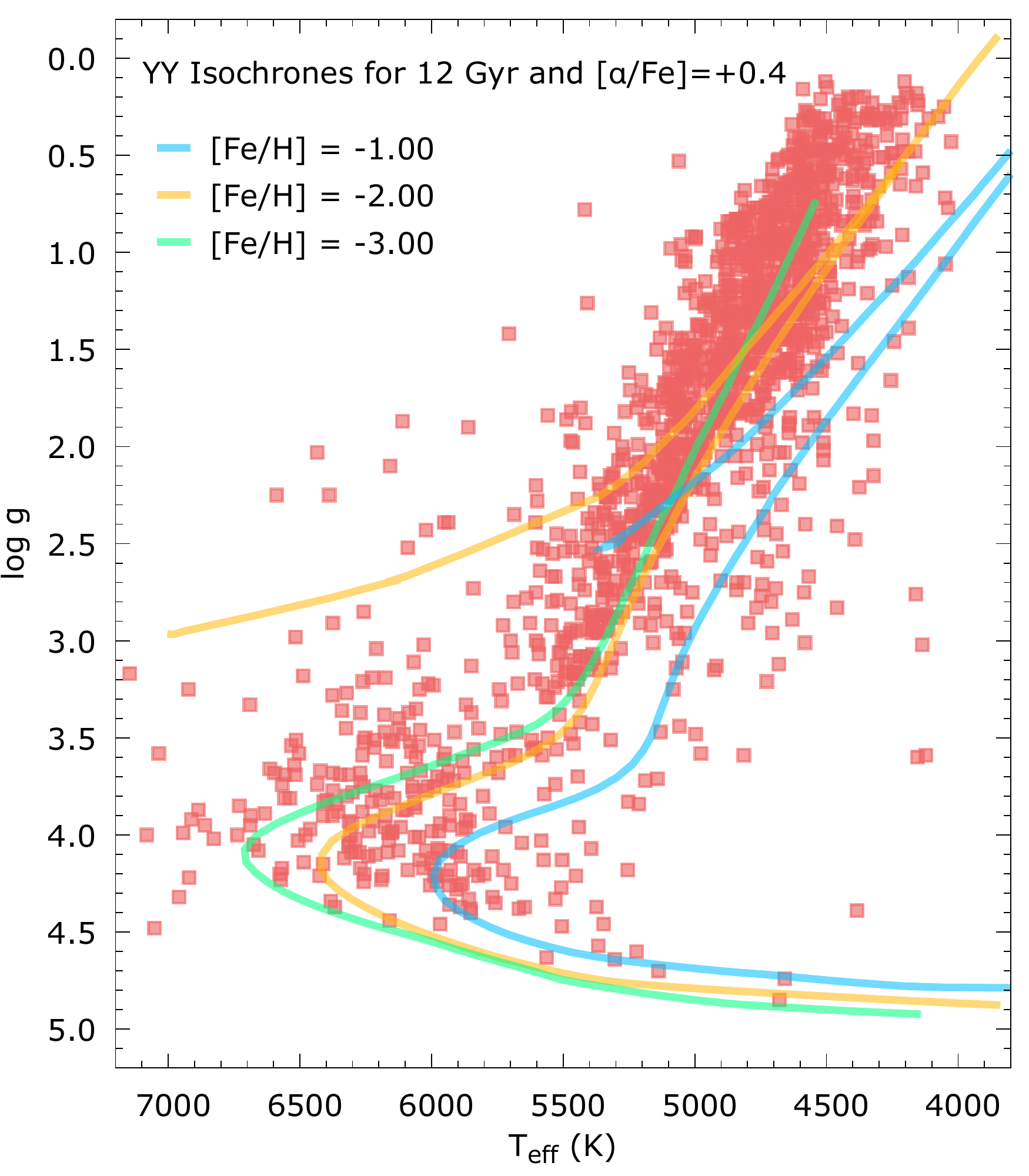}
\caption{H-R diagram for the program stars, using the parameters calculated by
the n-SSPP, listed in Table~\ref{comprave}. Overplotted are the YY Isochrones
\citep[12~Gy, 0.8~M$_{\rm \odot}$, \afe=$+$0.4;][]{demarque2004} for
\metal = $-$2.0, $-$2.5, and $-$3.0, and horizontal-branch tracks from
\citet{dotter2008}.}
\label{isochrone}
\end{figure}

From the 1,614 stars with estimated metallicities, 1,413 (88\%) have
\metal\,$\leq -1.0$, 980 (61\%) have \metal\,$\leq -2.0$, and 53 (3\%) of the
observed stars have \metal\,$\leq -3.0$.
The distribution of effective temperatures and surface gravities derived for the
RAVE follow-up sample are shown in Figure~\ref{isochrone}, compared with
Yale-Yonsei Isochrones \citep[12~Gy, 0.8~M$_{\rm \odot}$,
\afe=$+$0.4;][]{demarque2004} for \metal = $-$2.0, $-$2.5, and $-$3.0. Also
shown are the Horizontal-Branch tracks from the Dartmouth Stellar Evolution
Database \citep{dotter2008}, using the same input parameters. Since RAVE is a
magnitude-limited survey, it is expected that our sample would be dominated by
sub-giants and giants.
We also include corrections for the carbon abundances,
based on the stellar-evolution models presented in \citet{placco2014c}.
Typical uncertainties for the atmospheric parameters are 125\,K for \teff,
0.35\,dex for \logg, and 0.15-0.20\,dex for \metal, \cfe,\ and \afe.

\begin{figure*}[!ht]
\epsscale{0.97}
\plotone{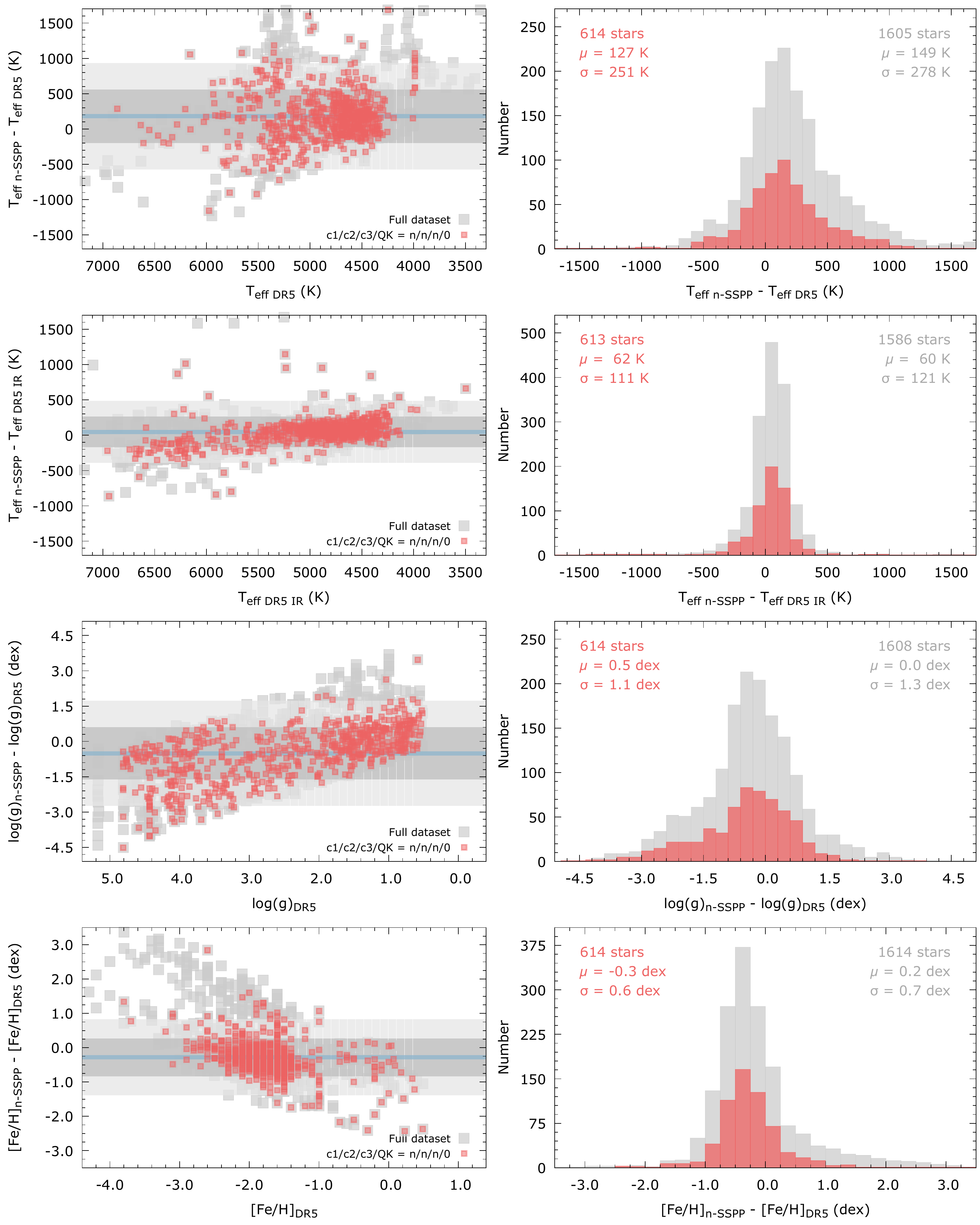}
\caption{Left panels: Differences between the atmospheric parameters determined
by the n-SSPP, $T_{\rm eff\, n-SSPP}$, logg$_{\rm\, n-SSPP}$, and [Fe/H]$_{\rm\,
n-SSPP}$, and the values from RAVE, $T_{\rm eff\, DR5}$, $T_{\rm eff\,DR5 IR}$,
logg$_{\rm\, DR5}$, and [Fe/H]$_{\rm\, DR5}$, reported by \citet{kunder2017}, as a
function of the RAVE spectroscopic values.
Filled symbols refer to the full sample of observed stars with determined
parameters (light-gray) and a subsample with
constraints on quality flags (red - see text for details). 
The horizontal solid line is the average of the residuals, while the
darker and lighter shaded areas represent the 1-$\sigma$ and 2-$\sigma$ regions,
respectively. Right panels: Histograms of the residuals between the 
n-SSPP and RAVE parameters shown in the left panels. Each panel also
lists the average offset and scatter determined from a Gaussian fit.}
\label{ravecomp} 
\end{figure*}

\begin{figure*}[!ht]
\epsscale{0.97}
\plotone{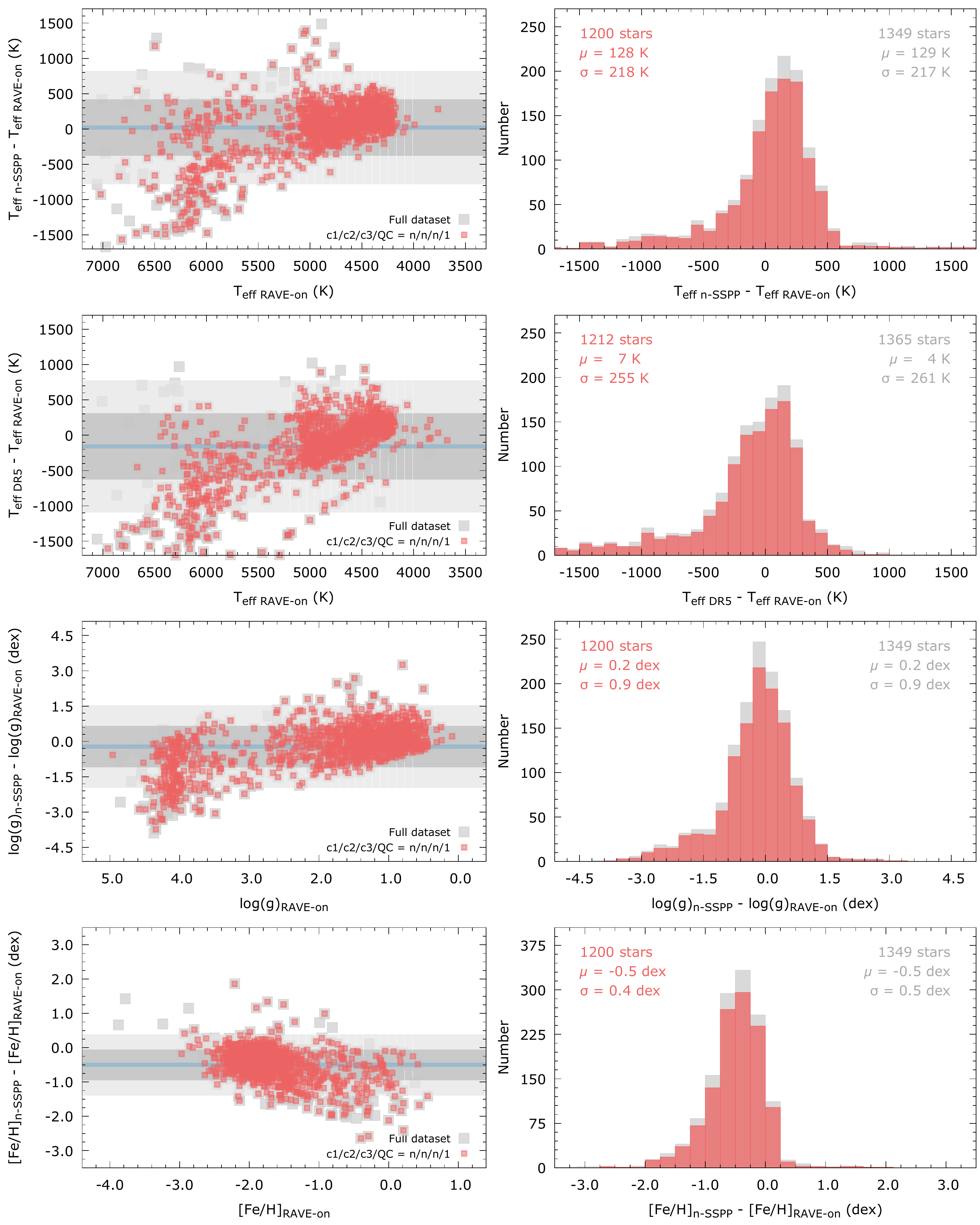}
\caption{Left panels: Differences between the atmospheric parameters determined
by the n-SSPP, $T_{\rm eff\, n-SSPP}$, logg$_{\rm\, n-SSPP}$, and
[Fe/H]$_{\rm\, n-SSPP}$, and the values from RAVE-on, $T_{\rm eff\,RAVE-on}$,
logg$_{\rm\, RAVE-on}$, and [Fe/H]$_{\rm\, RAVE-on}$, reported by
\citet{casey2017}, as a function of the RAVE-on spectroscopic values. 
Filled symbols refer to the full sample of observed stars with determined
parameters (light-gray) and a subsample with
constraints on quality flags (red - see text for details). 
The horizontal solid line is the average of
the residuals, while the darker and lighter shaded areas represent the
1-$\sigma$ and 2-$\sigma$ regions, respectively. Right panels: Histograms of
the residuals between the n-SSPP and RAVE parameters shown in the left panels.
Each panel also lists the average offset and scatter determined from a Gaussian
fit.}
\label{ravecompON} 
\end{figure*}

\section{Comparison with RAVE Parameters}
\label{secrave}

In this section we present a comparison between the atmospheric parameters
determined by the n-SSPP and the ones from the RAVE DR5 \citep{kunder2017} and
the RAVE-on pipelines \citep{casey2017}. The values used for these comparisons
(Figures~\ref{ravecomp} and \ref{ravecompON}) are listed in
Table~\ref{comprave}. Also shown in the table are the parameters from RAVE DR4
\citep{kordopatis2013}, from which the bulk of our target selection was made.
The RAVE pipeline derives parameters by a $\chi^2$ method using an extensive grid
of synthetic spectra \citep[see][for further details]{zwitter2008}.

\subsection{RAVE DR5}

The atmospheric parameters for RAVE DR5 were calculated with the DR4 stellar
pipeline, and calibrated using Kepler K2 seismic gravities, Gaia benchmark stars,
and results obtained from high-resolution studies \citep[see][for further
details]{kunder2017}. Thus, we refrain from using DR4 parameters for
the following analysis, but list them in Table~\ref{comprave} nonetheless.

Figure~\ref{ravecomp} presents the results of this comparison
for two different cases: (i) The full dataset, regardless of RAVE quality flag
values (light-gray filled squares), and (ii) The subsample of stars where the
RAVE pipeline converged and the first three
morphological flags indicate that the spectrum is of a normal star
({\texttt{QK == 0}} and {\texttt{c1/c2/c3 = n/n/n}} - red filled squares). 
The following discussion refers to the
comparison between the n-SSPP values and the subsample of RAVE stars with
parameters satisfying these criteria.

The left panels of Figure~\ref{ravecomp} show the differences between parameters
determined by the n-SSPP ($T_{\rm eff\,n-SSPP}$, \logg$_{\rm\, n-SSPP}$, and
[Fe/H]$_{\rm\, n-SSPP}$) and from RAVE (
$T_{\rm eff\,DR5}$ -- {\texttt{Teff\_N\_K}}, 
$T_{\rm eff\,DR5\,IR}$ -- {\texttt{Teff\_IR}},
\logg$_{\rm\, DR5}$ -- {\texttt{logg\_N\_K}}, and 
[Fe/H]$_{\rm\, DR5}$ -- {\texttt{Met\_N\_K}}), 
as a function of the RAVE DR5
spectroscopic values. Filled symbols refer to the stars observed as part of this
work. The horizontal solid line in each panel is the average of the residuals,
while the darker and lighter shaded areas represent the 1-$\sigma$ and
2-$\sigma$ regions, respectively. The right panels show histograms of the
residuals between the n-SSPP and RAVE parameters. Each panel also lists the
number of stars, the average offset, and the scatter determined from a
Gaussian fit to the residual distribution.

There are large deviations when comparing \teff\ values from RAVE and the
n-SSPP. Our determinations are consistently higher, in particular for \teff\,$<
4750$\,K. The zero-point offset on the residuals is 127\,K, and the scatter
is 251\,K. There is a somewhat good agreement between these estimates in the
[4500:5200]\,K range, and it is also possible to notice an upward trend on the
residuals for decreasing $T_{\rm eff\, DR5}$ values.
The RAVE DR5 catalog also provides temperature estimates based on the infrared flux
method of \citet{casagrande2010} ($T_{\rm eff\,DR5\,IR}$). These are in better agreement with the
n-SSPP values, showing a zero-point offset on the residuals of just 62\,K, and a
scatter of 111\,K.

The \logg\ comparison presents a more significant trend, with a zero-point offset on
the residuals of 0.5\,dex and a scatter of 1.1\,dex.

The behavior of the metallicity residuals follows similar trends as the \teff\
and \logg. This is expected, since the spectral features used for \metal\
estimates also change with temperature. The zero-point offset on the residual
distribution is $-$0.1\,dex and the scatter is 0.6\,dex.

\begin{figure*}[!ht]
\epsscale{1.175}
\plotone{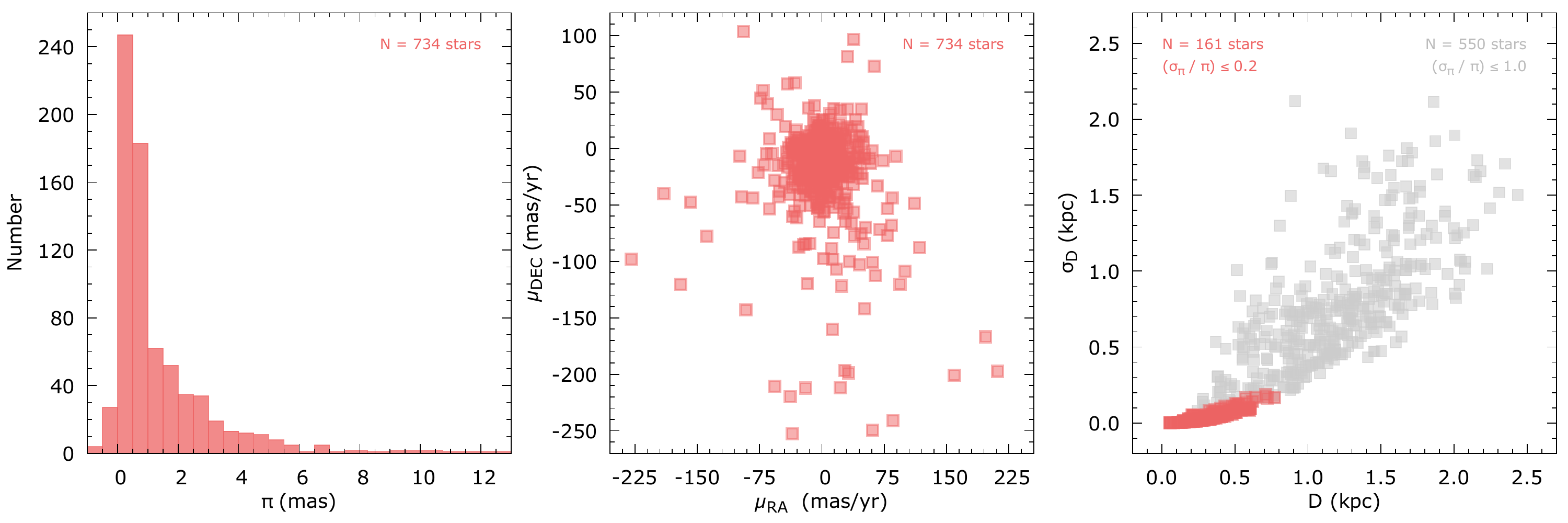}
\caption{Left panel: Parallax distribution for the observed RAVE stars found in
the Gaia TGAS catalog. Middle panel: Proper motions for these stars.  Right
panel: Derived  distances (with associated errors) from \citet{astra2016}, for
stars with $\sigma_\pi/\pi \leq 1.0$ (gray symbols) and $\sigma_\pi/\pi \leq
0.2$ (red symbols).}
\label{gaia}
\end{figure*}

\subsection{RAVE-on}

The RAVE-on catalog of stellar atmospheric parameters presents a re-analysis of
RAVE spectra using the data-driven code {\emph{The Cannon}} \citep{ness2015},
using data models from APOGEE and Kepler K2 \citep[see][for further
details]{casey2017}.
A comparison between the RAVE-on parameters and the n-SSPP parameters is presented in
Figure~\ref{ravecompON},
for the same cases shown above: (i) The full dataset, regardless of RAVE-on
quality flag values (light-gray filled squares), and (ii) The subsample of
stars where the RAVE-on pipeline converged and the
first three morphological flags indicate that the spectrum is of a normal
star ({\texttt{QC == 1}} and {\texttt{c1/c2/c3 = n/n/n}} - red filled squares).

The left panels of Figure~\ref{ravecompON} show the differences between parameters
determined by the n-SSPP ($T_{\rm eff\, n-SSPP}$, \logg$_{\rm\, n-SSPP}$, and
[Fe/H]$_{\rm\, n-SSPP}$) and from RAVE-on ($T_{\rm eff\,RAVE-on}$, \logg$_{\rm\,
RAVE-on}$, and [Fe/H]$_{\rm\, RAVE-on}$), as a function of the RAVE-on values.
Also shown is the comparison between the effective temperatures in RAVE DR5 and
RAVE-on. Filled symbols refer to the stars observed as part of this work. The
horizontal solid line in each panel is the average of the residuals, while the
darker and lighter shaded areas represent the 1-$\sigma$ and 2-$\sigma$ regions,
respectively. The right panels show histograms of the residuals between the
n-SSPP and RAVE parameters. Each panel also lists the average offset and scatter
determined from a Gaussian fit to the residual distribution.

The deviations, when comparing \teff\ values from RAVE-on and the
n-SSPP are large, in particular for \teff\,$> 5000$\,K. 
The zero-point offset in the residuals and the scatter (128\,K and 218\,K,
respectively) are slightly smaller than RAVE DR5 vs. n-SSPP, as seen in
Figure~\ref{ravecomp}. 
For \teff\,$< 5000$\,K, the spread between the determinations is smaller, but
there is a noticeable offset, with the n-SSPP temperatures being higher
when compared to RAVE-on.
We also compared the RAVE-on temperatures with the values from RAVE DR5.  The
correlation is worse, even though the  zero-point offset in the residuals
is just 7\,K, and the scatter is 255\,K. Still, similar to the other temperature
comparisons, the agreement is somewhat better in the [4500:5200]\,K range.

The \logg\ comparison exhibits better agreement than the one between the n-SSPP and
RAVE DR5, with a zero-point offset of 0.2\,dex and a scatter of 0.9\,dex.
The scatter is mostly driven by values with \logg\,$_{\rm RAVE-on} \gtrsim
3.0$.

The behavior of the metallicity residuals follows the same trends as the \teff\
and \logg. The agreement similar to the comparison between the n-SSPP and the
RAVE DR5 values shown in Figure~\ref{ravecomp}. The zero-point offset in the
residual distribution is $-$0.5\,dex and the scatter is 0.4\,dex. The n-SSPP
values are consistently smaller than the RAVE-on determinations, with a larger
scatter for [Fe/H]$_{\rm RAVE-on} \gtrsim -1.0$.

\begin{figure*}[!ht]
\epsscale{2.30}
\plottwo{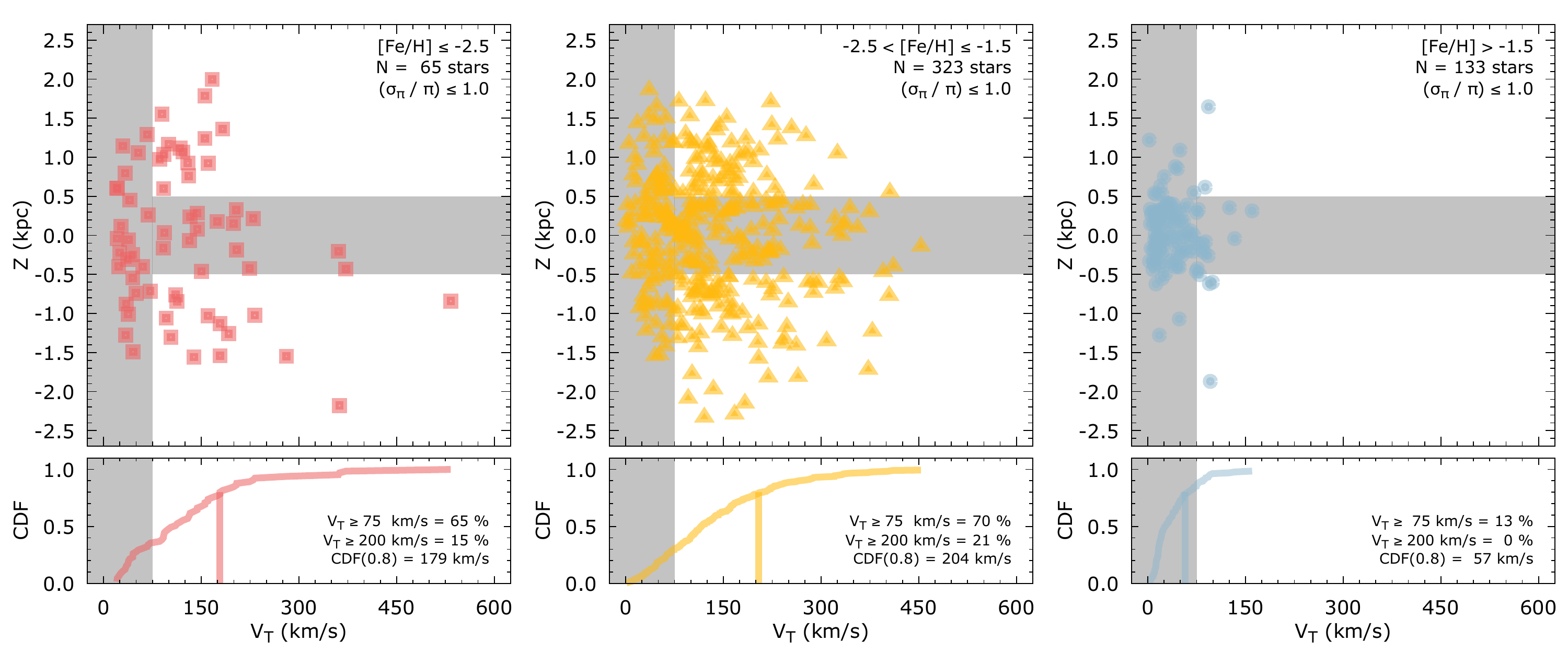}{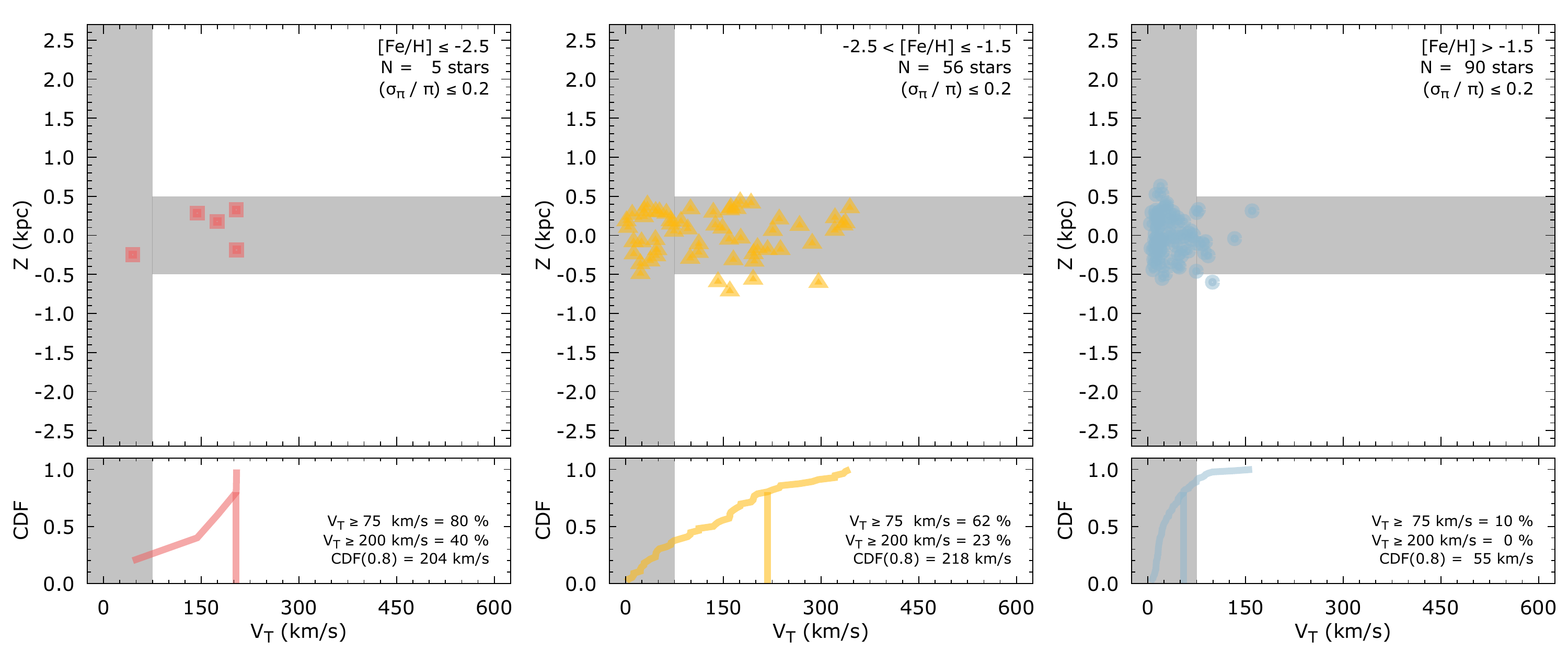}

\caption{Distance from the Galactic plane (Z, in kpc), as a function of the
transverse velocity (V$_{\rm T}$, in ${\rm km\,s^{-1}}$), for three metallicity
regimes. The horizontal shaded area roughly marks the region potentially
occupied by thin-disk stars in the Galaxy ($\pm 500\,{\rm pc}$), and the
vertical shaded area represents typical transverse velocities occupied by
thin-disk stars (see text for details).  Also shown are the V$_{\rm T}$
cumulative distribution functions for the \metal\ intervals. The values shown
in each panel represent the fraction of stars for V$_{\rm T} \geq 75\,{\rm
km\,s}^{-1}$, and V$_{\rm T} \geq 200\,{\rm km\, s}^{-1}$, and the V$_{\rm T}$
value where the CDF reaches a fraction of 80\%. The upper panels show the stars
with accepted distances from TGAS ($\sigma_\pi/\pi \leq 1.0$), and the lower
panel shows a more restricted error cut on parallaxes ($\sigma_\pi/\pi \leq
0.2$).}

\label{vtz} \end{figure*}

\section{Sample stars observed with Gaia}
\label{gaias}

We searched the Tycho-Gaia Astrometric Solution database
\citep[TGAS;][]{gaia2016,lindegren2016}, and found 734 matches with stars in
our sample. Figure~\ref{gaia} (left and middle panels) shows the distribution
of the parallaxes and proper motions for these stars.
Distances (and associated errors) were taken from the catalog of
\citet{astra2016}, who infered distances for the TGAS objects using not only
observed parallaxes and their uncertainties, but also an anisotropic prior
derived from the observability of stars in a Milky Way model \citep[see][for
further details]{bailer2015}. The right panel of Figure~\ref{gaia} shows
distances (and associated errors) for stars with $\sigma_{\pi}/\pi \leq 0.2$
and $\sigma_{\pi}/\pi \leq 1.0$.  Although the later is a fairly relaxed
constraint, we chose it in order to have a larger sample of stars to exemplify
our improved selection criteria (see below).  Because of that, distances are
limited to $D \leq 2.5\,{\rm kpc}$. 
Table~\ref{gaiar} lists, for the 734 stars in common with TGAS, the Gaia ID,
$G$ magnitude, proper motions ($\mu_{{\rm RA}}$ and $\mu_{{\rm DEC}}$), and
parallaxes ($\pi$). Distances and errors infered from the Milky Way prior were
taken from the \citet{astra2016} catalog.

\begin{figure*}[!ht]
\epsscale{1.16}
\plottwo{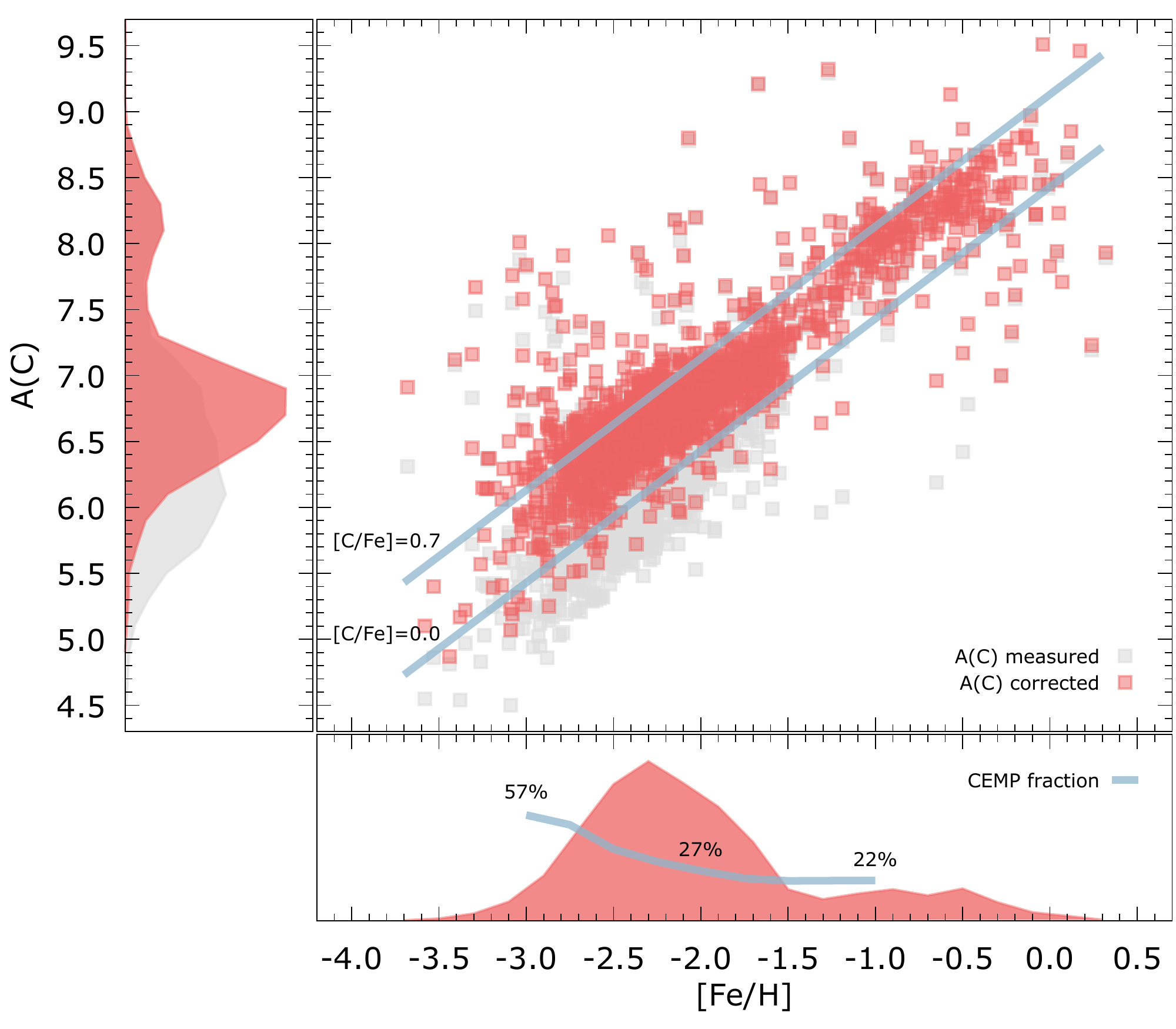}{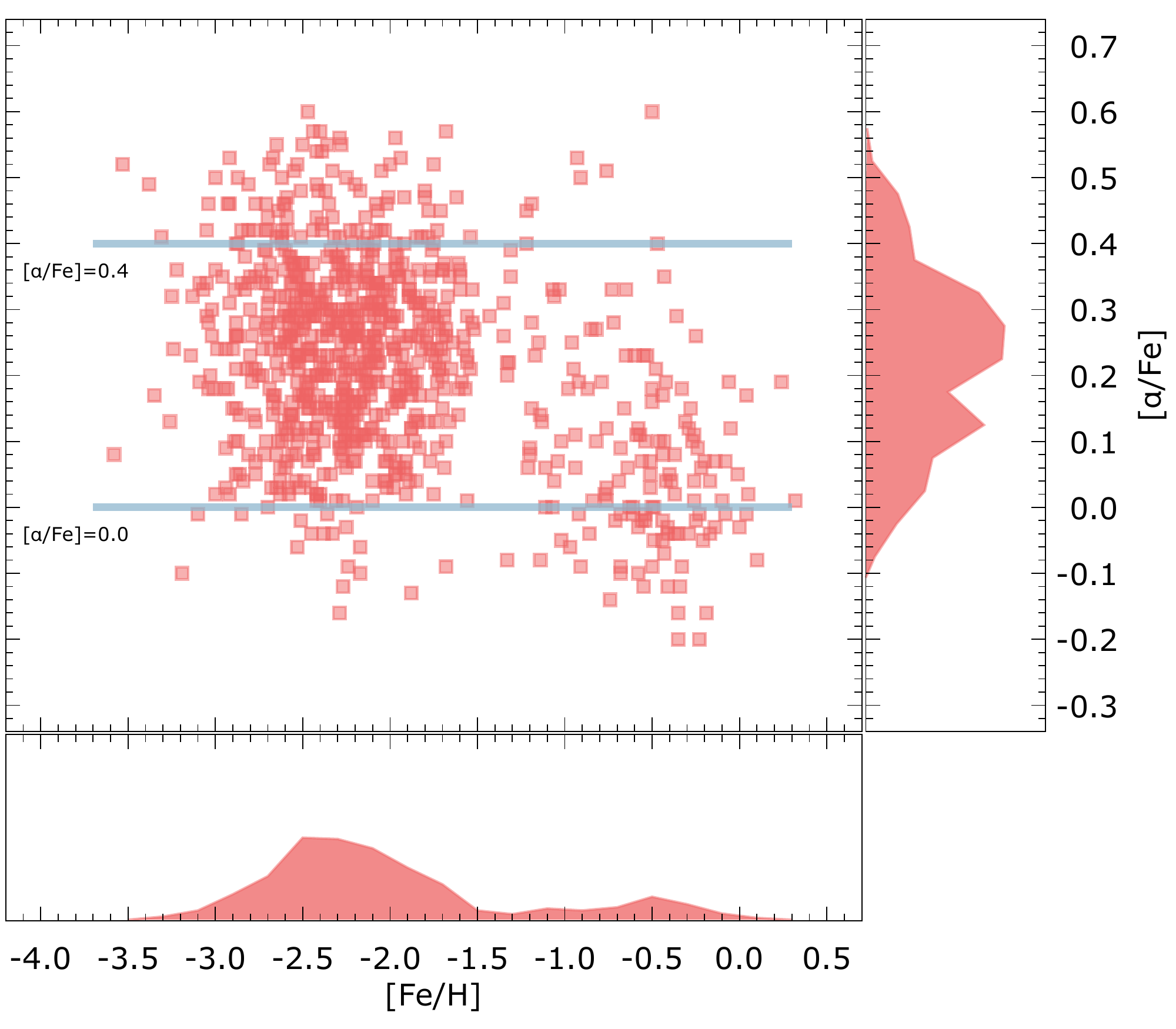}
\caption{Absolute carbon ($A$(C), measured and corrected - left panel), and relative
$\alpha$-element abundances (right panel), as a function of the metallicity
calculated by the n-SSPP. The side and lower panels show the marginalized
distributions for each quantity. The solid line in the lower panel shows the
cumulative CEMP fractions for the stars with $-3.0 \leq$\,\metal\,$\leq -1.0$.}
\label{acfeh} 
\end{figure*}

Having kinematic information for these stars in addition to atmospheric
parameters is a valuable tool for better pre-selecting low-metallicity stars
for spectroscopic follow-up.
Using the information provided by TGAS, we were able to calculate the Cartesian
coordinates and transverse velocities for the sample with $\sigma_{\pi}/\pi \leq 1.0$.
The upper panels of Figure~\ref{vtz} show the behavior of the height above the
Galactic plane (Z, in kpc) as a function of the transverse velocity (V$_{\rm
T}$, in ${\rm km\, s}^{-1}$), for three metallicity regimes. The shaded areas
roughly mark the region occupied by thin-disk stars \citep[$\pm 500\,{\rm
pc}$;][]{recio-blanco2014} and typical transverse velocities for thin-disk
stars.  Also shown are the V$_{\rm T}$ cumulative distribution functions (CDF).
Values on each panel represent the fraction of stars in each \metal\ regime
with V$_{\rm T} \geq 75\,{\rm km\, s}^{-1}$ and V$_{\rm T} \geq 200\,{\rm km\,
s}^{-1}$, and also the V$_{\rm T}$ value where the CDF reaches a fraction of
80\%.  The bottom panels of Figure~\ref{vtz} show the same quantities, using a
more restrictive error cut on parallaxes ($\sigma_{\pi}/\pi \leq 0.2$; 151
stars).

From inspection of Figure~\ref{vtz}, it can be seen that the transverse
velocity is an effective additional constraint to select low-metallicity stars.
The CDF for the high-metallicity regime reaches 80\% at V$_{\rm T} = 57\,{\rm
km\,s}^{-1}$, while the more metal-poor regimes only reach 80\% at V$_{\rm T}
\geq 180\,{\rm km\, s}^{-1}$. In fact, at V$_{\rm T} \geq 200\,{\rm
km\,s}^{-1}$, there are no stars in our sample with \metal$ > -1.5$. The same
applies to the distance-limited sample shown on the lower panels of
Figure~\ref{vtz}. By application of a search criteria in V$_{\rm T}$ (but not
in Z), one can successfully search for stars belonging to the halo population,
but currently located close to the Galactic plane.

To roughly quantify the improvement in the search for low-metallicity stars
using the transverse velocity, we use the V$_{\rm T}$ and \metal\ for the 550
stars shown in the right panel of Figure~\ref{gaia}. Within those, 82\% have
\metal$ \leq -1.0$ and 50\% have \metal$ \leq -2.0$. By selecting stars with
V$_{\rm T} \geq 75\,{\rm km/s}$ and V$_{\rm T} \geq 200\,{\rm km/s}$, these
fractions would increase to 97\%/58\% and 100\%/60\%, respectively, albeit with
the disadvantage that a kinematic bias is injected into the resulting sample.
We caution the reader that the present sample {\emph{was chosen}} to include
only low-metallicity stars based on RAVE parameters. Nonetheless, the
quantifiable improvements in the fractions suggest that these constraints
should also be robust in non-\metal-biased samples, and can be used to tailor
future searches for metal-poor stars.

\section{Carbon and $\alpha$-element Abundances}
\label{seccomp}

In a sample of low-metallicity stars, one should expect an increasing fraction
of CEMP (\metal\,$\leq -1.0$ and \cfe\,$\geq +0.7$) stars for decreasing
metallicities \citep{rossi1999,rossi2005,beers2005,placco2010,placco2011}.
Among the 1,606 stars observed in this work for which carbon abundances could
be determined, we identified 106 CEMP stars, before applying the carbon corrections of
\citet{placco2014c}, and 306 CEMP stars, after corrections were applied. Since
RAVE does not provide carbon abundances, we have the opportunity to add yet
another selection criteria when assembling target lists for high-resolution
spectroscopic follow-up.

Figure~\ref{acfeh} shows, on the left side, the absolute carbon abundances
($A(C)$\footnote{$A(C) = log(N_C/{}N_H) + 12$}, measured and corrected), as a
function of the metallicity calculated by the n-SSPP. The side and lower panels
show the marginalized distributions for each quantity.  The right side of
Figure~\ref{acfeh} shows the $\alpha$-element abundance ratios as a function of
the metallicity.  The \afe\ ratio behaves similarly to the sample of
\citet{cayrel2004}, with values typically ranging from 0.0 to $+0.4$.

The sample observed in this work also provides for an independent (and
unbiased) calculation of the fractions of carbon-enhanced stars as a function of
metallicity.  Results are shown in the bottom-left panel of Figure~\ref{acfeh},
where the solid line represents the cumulative CEMP fractions for stars with
$-3.0 \leq$\,\metal\,$\leq -1.0$. The fraction for \metal\,$\leq -2.0$ ($27 \pm
3\%$\footnote{Uncertainties in the fractions represent the Wilson score
confidence intervals \citep{wilson1927}. See \citet{yoon2018} for further
details.}) agrees within uncertainties with the $20 \pm 13\%$ fraction reported
by \citet{placco2014c}, but is somewhat higher than the $13 \pm 1\%$ fraction
reported by \citet{lee2013}. For \metal\,$\leq -3.0$, the fraction found in this
work ($57 ^{+13} _{-14}\%$) is higher than both Placco et al. ($43 \pm 5\%$) and
Lee et al.  ($23 \pm 3\%$).

The CEMP stars identified in this work can be further sub-classified, based on
their position in the $Yoon-Beers$ A(C) vs. \metal\ diagram, following
\citet{yoon2016}. Roughly, one would expect to find (assuming \metal\,$\leq
-1.0$ and \cfe\,$\geq +0.7$) CEMP-$s$ (or CEMP-$i$) stars for $-3.5
\lesssim$\,\metal\,$ \lesssim -2.5$ and A(C)\,$ \gtrsim 7.25$ or \metal\,$
\gtrsim -2.5$ (Group I); and CEMP-no stars for \metal\,$ \lesssim -2.5$, A(C)\,$
\lesssim 7.25$, and \cfe\,$ \leq +1.5$ (Group II) or \cfe\,$ > +1.5$ (Group
III). Based on these, there are 169 CEMP Group~I, 131 CEMP Group~II, and 6 CEMP
Group~III stars. 
Table~\ref{groups} lists the main parameters for all 306 CEMP stars, including
the assignment between Groups I/II/III.

\begin{figure}[!ht]
\epsscale{1.16}
\plotone{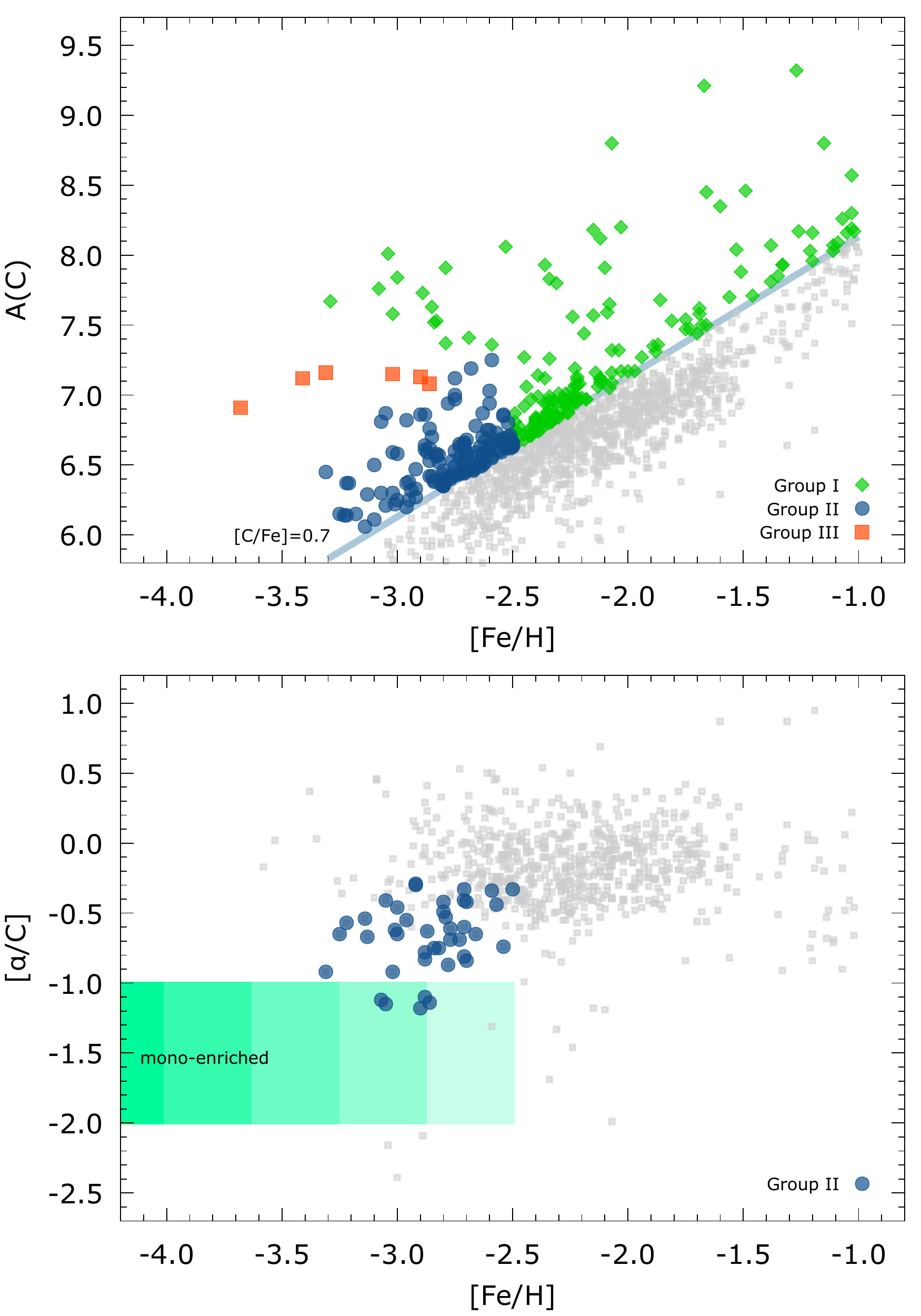}
\caption{Upper panel: $A$(C), as a function of the n-SSPP metallicity, for stars
with \protect\metal$ \leq -1.0$. CEMP stars belonging to Groups I, II, and III are
identified, following the criteria described in the text. Lower panel:
\protect\abund{\alpha}{C} abundance ratio. The shaded area outlines the region
where mono-enriched second-generation stars would preferably be found.}
\label{acmgc} 
\end{figure}

The upper panel of
Figure~\ref{acmgc} shows the distribution of CEMP stars among these groups. 
We note that these limits and classifications aim to provide a first-step criteria
for selection, and the region at \metal\,$ \sim -2.5$ is where the groups in
\citet{yoon2016} overlap.  The enhancement in carbon in Group~I stars is
the result of external pollution from an AGB companion in a binary system.
Because of that, these objects are not suitable to probe the chemical
evolution of the Galaxy.  Groups~II and III, in contrast, are believed to
describe stars formed from gas clouds that were enriched by early SNe events from
massive stars. 

\citet{hartwig2018} argue that that the so-called {\emph{mono-enriched}} stars
(indicating a single first-star progenitor) can be found at metallicities as
high as \metal\,$\sim -2.0$, and can be separated from other
{\emph{multiple-enriched}} stars, based on the \abund{Mg}{C} ratio. Since the
n-SSPP calculates \afe, in principle, the \abund{\alpha}{C} ratio can be used to
identify true mono-enriched stars. The lower panel of Figure~\ref{acmgc} shows
the distribution of this quantity for the stars in the sample, highlighting the
CEMP Group~II stars.  Note that all six CEMP Group~III were observed with
ESO/NTT, hence there are no \afe\, determinations available for these stars.
The shaded area outlines the region where mono-enriched second-generation stars
would preferably be found \citep{hartwig2018}.

\section{Conclusions}
\label{final}

We have presented results from a medium-resolution ($R \sim 1,200 - 2,000$)
spectroscopic follow-up of low-metallicity stars selected from the RAVE
database. Our observing campaign ran from semester 2014A to 2017A, and used six
different telescope/instrument configurations, in both the Southern and Northern
Hemispheres.  Atmospheric parameters and abundances for carbon and the
$\alpha$-elements were calculated using our well-tested n-SSPP pipeline. From
the 1,694 unique stars observed, 1,413 were confirmed to be metal-poor
(\metal\,$\leq -1.0$), and 306 were carbon-enhanced (\cfe\,$ \geq +0.7$), after
evolutionary corrections have been applied.

Comparison with atmospheric-parameter estimates from RAVE DR5 and RAVE-on
revealed discrepancies, in particular for temperatures, with a 127\,K zero-point
offset in the residual distribution, and a 251\,K scatter (n-SSPP vs. RAVE DR5).
These differences led to similar inconsistencies in \logg\, and \metal,
resulting in 202 stars with \metal$_{\rm\, n-SSPP} \geq -1.0$ being included in
our sample. The agreement is somewhat better between the RAVE-on values and the
n-SSPP, however with large scatter for \teff$_{\rm\, RAVE-on} \gtrsim 5000$\,K
and \logg$_{\rm\, RAVE-on} \gtrsim 3.0$. A forthcoming paper will compare the
n-SSPP determinations with values from ongoing high-resolution spectroscopic
follow-up by the R-Process Alliance; then we will be able to better assess the
nature of the differences presented in this work.

A search in the Gaia/TGAS database revealed matches for 734 stars observed in
this work. Based on a height above the Galactic plane vs. transverse velocity
diagram, we show that these kinematical properties can be further used to
pre-select low-metallicity stars as well, increasing the success rate of similar
spectroscopic follow-ups in the future.

The atmospheric parameters and carbon abundances determined by this work not
only characterize the sample of halo stars but also serve as a stepping stone
for constructing target lists for a variety of follow-up studies
\citep[e.g.][]{hansen2018}. In this regard, the most metal-poor stars
identified (including the CEMP Group~III and the {\emph{mono-enriched
stars}}) have already been followed-up with high-resolution spectroscopy for
detailed abundance studies. One $r$-II star with detected uranium
\citep{placco2017}, a bright $r$-II star at \metal$\sim -2$ \citep{sakari2018},
and the first $r+s$ star 
\citep[showing the {\emph{combined}} signature of the $r$-process and
$s$-process;][]{gull2018}
have also been found.  Furthermore, \citet{placco2016b} used an UMP star from
RAVE to help constrain the mass distribution of the first stars in the
universe. This work is ongoing, with more publications to follow in due course.
Overall, these efforts will continue to expand our understanding of the
nucleosynthesis of the elements and early chemical evolution, which forms the
basis for understanding star- and galaxy-formation in the early universe.

%

\acknowledgments
The authors acknowledge partial support for this work from grant PHY 14-30152;
Physics Frontier Center/JINA Center for the Evolution of the Elements
(JINA-CEE), awarded by the US National Science Foundation.  
S.R. acknowledges partial support from FAPESP, CAPES/PROEX, and CNPq.
Y.S.L. acknowledges support from the National Research Foundation (NFR) of Korea
grant funded by the Ministry of Science and ICT (No.2017R1A5A1070354,
NRF-2015R1C1A1A02036658, and NRF-2018R1A2B6003961).
%
%
Based on observations at Kitt Peak National Observatory, National Optical
Astronomy Observatory (NOAO Prop. IDs: 14A-0323, 14B-0231, 15A-0071, 15B-0071,
17A-0295), which is operated by the Association of Universities for Research in
Astronomy (AURA) under cooperative agreement with the National Science
Foundation. The authors are honored to be permitted to conduct astronomical
research on Iolkam Du'ag (Kitt Peak), a mountain with particular significance
to the Tohono O'odham. 
%
%
Based on observations obtained at the Gemini Observatory (Prop. IDs:
GN-2015A-Q-401, GS-2015A-Q-205), which is operated by the Association of
Universities for Research in Astronomy, Inc., under a cooperative agreement
with the NSF on behalf of the Gemini partnership: the National Science
Foundation (United States), the National Research Council (Canada), CONICYT
(Chile), Ministerio de Ciencia, Tecnolog\'{i}a e Innovaci\'{o}n Productiva
(Argentina), and Minist\'{e}rio da Ci\^{e}ncia, Tecnologia e Inova\c{c}\~{a}o
(Brazil).
%
%
Based on observations obtained at the Southern Astrophysical Research (SOAR)
telescope (Prop. IDs: 2014B-0231, 2015A-0071, 2015B-0020, 2016A-0019,
2017A-0016), which is a joint project of the Minist\'{e}rio da Ci\^{e}ncia,
Tecnologia, e Inova\c{c}\~{a}o (MCTI) da Rep\'{u}blica Federativa do Brasil,
the U.S. National Optical Astronomy Observatory (NOAO), the University of North
Carolina at Chapel Hill (UNC), and Michigan State University (MSU).
%
%
Based on observations collected at the European Organisation for Astronomical
Research in the Southern Hemisphere under ESO programme(s) 095.D-0202(A),
096.D-0018(A), 097.D-0196(A), 098.D-0434(A), and 099.D-0428(A).
This work has made use of data from the European Space Agency (ESA) mission
{\it Gaia} (\url{https://www.cosmos.esa.int/gaia}), processed by the {\it Gaia}
Data Processing and Analysis Consortium (DPAC,
\url{https://www.cosmos.esa.int/web/gaia/dpac/consortium}). Funding for the
DPAC has been provided by national institutions, in particular the institutions
participating in the {\it Gaia} Multilateral Agreement.
This research has made use of NASA's Astrophysics Data System Bibliographic
Services; the arXiv pre-print server operated by Cornell University; the SIMBAD
database hosted by the Strasbourg Astronomical Data Center; and the online Q\&A
platform {\texttt{stackoverflow}}
(\href{http://stackoverflow.com/}{http://stackoverflow.com/}).

\software{
{\texttt{awk}}\,\citep{awk}, 
{\texttt{gnuplot}}\,\citep{gnuplot}, 
{\texttt{IRAF}}\,\citep{tody1986,tody1993}, 
{\texttt{n-SSPP}}\,\citep{beers2014}, 
{\texttt{R-project}}\,\citep{rproject}, 
{\texttt{sed}}\,\citep{sed}.
}

\bibliographystyle{apj}

\begin{thebibliography}{}
\expandafter\ifx\csname natexlab\endcsname\relax\def\natexlab#1{#1}\fi

\bibitem[{{Abbott} {et~al.}(2017){Abbott}, {Abbott}, {Abbott}, {Acernese},
  {Ackley}, {Adams}, {Adams}, {Addesso}, {Adhikari}, {Adya}, \&
  et~al.}]{abbott2017}
{Abbott}, B.~P., {Abbott}, R., {Abbott}, T.~D., {et~al.} 2017, \apjl, 848, L12

\bibitem[{Aho {et~al.}(1987)Aho, Kernighan, \& Weinberger}]{awk}
Aho, A.~V., Kernighan, B.~W., \& Weinberger, P.~J. 1987, The AWK Programming
  Language (Boston, MA, USA: Addison-Wesley Longman Publishing Co., Inc.)

\bibitem[{{Aoki} {et~al.}(2007){Aoki}, {Beers}, {Christlieb}, {Norris}, {Ryan},
  \& {Tsangarides}}]{aoki2007}
{Aoki}, W., {Beers}, T.~C., {Christlieb}, N., {et~al.} 2007, \apj, 655, 492

\bibitem[{{Astraatmadja} \& {Bailer-Jones}(2016)}]{astra2016}
{Astraatmadja}, T.~L., \& {Bailer-Jones}, C.~A.~L. 2016, \apj, 833, 119

\bibitem[{{Bailer-Jones}(2015)}]{bailer2015}
{Bailer-Jones}, C.~A.~L. 2015, \pasp, 127, 994

\bibitem[{{Barklem} {et~al.}(2005){Barklem}, {Christlieb}, {Beers}, {Hill},
  {Bessell}, {Holmberg}, {Marsteller}, {Rossi}, {Zickgraf}, \&
  {Reimers}}]{barklem2005}
{Barklem}, P.~S., {Christlieb}, N., {Beers}, T.~C., {et~al.} 2005, \aap, 439,
  129

\bibitem[{{Beers} \& {Christlieb}(2005)}]{beers2005}
{Beers}, T.~C., \& {Christlieb}, N. 2005, \araa, 43, 531

\bibitem[{{Beers} {et~al.}(2014){Beers}, {Norris}, {Placco}, {Lee}, {Rossi},
  {Carollo}, \& {Masseron}}]{beers2014}
{Beers}, T.~C., {Norris}, J.~E., {Placco}, V.~M., {et~al.} 2014, \apj, 794, 58

\bibitem[{{Beers} {et~al.}(1985){Beers}, {Preston}, \& {Shectman}}]{beers1985}
{Beers}, T.~C., {Preston}, G.~W., \& {Shectman}, S.~A. 1985, \aj, 90, 2089

\bibitem[{{Beers} {et~al.}(1992){Beers}, {Preston}, \& {Shectman}}]{beers1992}
---. 1992, \aj, 103, 1987

\bibitem[{{Beers} {et~al.}(2017){Beers}, {Placco}, {Carollo}, {Rossi}, {Lee},
  {Frebel}, {Norris}, {Dietz}, \& {Masseron}}]{beers2017}
{Beers}, T.~C., {Placco}, V.~M., {Carollo}, D., {et~al.} 2017, \apj, 835, 81

\bibitem[{{Bessell} {et~al.}(2015){Bessell}, {Collett}, {Keller}, {Frebel},
  {Heger}, {Casey}, {Masseron}, {Asplund}, {Jacobson}, {Lind}, {Marino},
  {Norris}, {Yong}, {Da Costa}, {Chan}, {Magic}, {Schmidt}, \&
  {Tisserand}}]{bessell2015}
{Bessell}, M., {Collett}, R., {Keller}, S., {et~al.} 2015, ArXiv e-prints,
  arXiv:1505.03756

\bibitem[{{Burbidge} {et~al.}(1957){Burbidge}, {Burbidge}, {Fowler}, \&
  {Hoyle}}]{b2fh}
{Burbidge}, E.~M., {Burbidge}, G.~R., {Fowler}, W.~A., \& {Hoyle}, F. 1957,
  Reviews of Modern Physics, 29, 547

\bibitem[{{Cameron}(1957)}]{cameron1957}
{Cameron}, A.~G.~W. 1957, \pasp, 69, 201

\bibitem[{{Casagrande} {et~al.}(2010){Casagrande}, {Ram{\'{\i}}rez},
  {Mel{\'e}ndez}, {Bessell}, \& {Asplund}}]{casagrande2010}
{Casagrande}, L., {Ram{\'{\i}}rez}, I., {Mel{\'e}ndez}, J., {Bessell}, M., \&
  {Asplund}, M. 2010, \aap, 512, A54

\bibitem[{{Casey} {et~al.}(2017){Casey}, {Hawkins}, {Hogg}, {Ness}, {Rix},
  {Kordopatis}, {Kunder}, {Steinmetz}, {Koposov}, {Enke}, {Sanders}, {Gilmore},
  {Zwitter}, {Freeman}, {Casagrande}, {Matijevi{\v c}}, {Seabroke},
  {Bienaym{\'e}}, {Bland-Hawthorn}, {Gibson}, {Grebel}, {Helmi}, {Munari},
  {Navarro}, {Reid}, {Siebert}, \& {Wyse}}]{casey2017}
{Casey}, A.~R., {Hawkins}, K., {Hogg}, D.~W., {et~al.} 2017, \apj, 840, 59

\bibitem[{{Cayrel} {et~al.}(2004){Cayrel}, {Depagne}, {Spite}, {Hill}, {Spite},
  {Fran{\c c}ois}, {Plez}, {Beers}, {Primas}, {Andersen}, {Barbuy},
  {Bonifacio}, {Molaro}, \& {Nordstr{\"o}m}}]{cayrel2004}
{Cayrel}, R., {Depagne}, E., {Spite}, M., {et~al.} 2004, \aap, 416, 1117

\bibitem[{{Chiaki} {et~al.}(2017){Chiaki}, {Tominaga}, \&
  {Nozawa}}]{chiaki2017}
{Chiaki}, G., {Tominaga}, N., \& {Nozawa}, T. 2017, \mnras, 472, L115

\bibitem[{{Christlieb} {et~al.}(2008){Christlieb}, {Sch{\"o}rck}, {Frebel},
  {Beers}, {Wisotzki}, \& {Reimers}}]{christlieb2008}
{Christlieb}, N., {Sch{\"o}rck}, T., {Frebel}, A., {et~al.} 2008, \aap, 484,
  721

\bibitem[{{Christlieb} {et~al.}(2002){Christlieb}, {Bessell}, {Beers},
  {Gustafsson}, {Korn}, {Barklem}, {Karlsson}, {Mizuno-Wiedner}, \&
  {Rossi}}]{christlieb2002}
{Christlieb}, N., {Bessell}, M.~S., {Beers}, T.~C., {et~al.} 2002, \nat, 419,
  904

\bibitem[{{Christlieb} {et~al.}(2004){Christlieb}, {Beers}, {Barklem},
  {Bessell}, {Hill}, {Holmberg}, {Korn}, {Marsteller}, {Mashonkina}, {Qian},
  {Rossi}, {Wasserburg}, {Zickgraf}, {Kratz}, {Nordstr{\"o}m}, {Pfeiffer},
  {Rhee}, \& {Ryan}}]{christlieb2004}
{Christlieb}, N., {Beers}, T.~C., {Barklem}, P.~S., {et~al.} 2004, \aap, 428,
  1027

\bibitem[{{Cooke} \& {Madau}(2014)}]{cooke2014}
{Cooke}, R., \& {Madau}, P. 2014, ArXiv e-prints, arXiv:1405.7369

\bibitem[{{Cooke} {et~al.}(2012){Cooke}, {Pettini}, \& {Murphy}}]{cooke2012}
{Cooke}, R., {Pettini}, M., \& {Murphy}, M.~T. 2012, \mnras, 425, 347

\bibitem[{{Demarque} {et~al.}(2004){Demarque}, {Woo}, {Kim}, \&
  {Yi}}]{demarque2004}
{Demarque}, P., {Woo}, J.-H., {Kim}, Y.-C., \& {Yi}, S.~K. 2004, \apjs, 155,
  667

\bibitem[{{Dotter} {et~al.}(2008){Dotter}, {Chaboyer}, {Jevremovi{\'c}},
  {Kostov}, {Baron}, \& {Ferguson}}]{dotter2008}
{Dotter}, A., {Chaboyer}, B., {Jevremovi{\'c}}, D., {et~al.} 2008, \apjs, 178,
  89

\bibitem[{{Drout} {et~al.}(2017){Drout}, {Piro}, {Shappee}, {Kilpatrick},
  {Simon}, {Contreras}, {Coulter}, {Foley}, {Siebert}, {Morrell}, {Boutsia},
  {Di Mille}, {Holoien}, {Kasen}, {Kollmeier}, {Madore}, {Monson},
  {Murguia-Berthier}, {Pan}, {Prochaska}, {Ramirez-Ruiz}, {Rest}, {Adams},
  {Alatalo}, {Ba{\~n}ados}, {Baughman}, {Beers}, {Bernstein}, {Bitsakis},
  {Campillay}, {Hansen}, {Higgs}, {Ji}, {Maravelias}, {Marshall}, {Bidin},
  {Prieto}, {Rasmussen}, {Rojas-Bravo}, {Strom}, {Ulloa},
  {Vargas-Gonz{\'a}lez}, {Wan}, \& {Whitten}}]{drout2017}
{Drout}, M.~R., {Piro}, A.~L., {Shappee}, B.~J., {et~al.} 2017, Science, 358,
  1570

\bibitem[{{Frebel} {et~al.}(2015){Frebel}, {Chiti}, {Ji}, {Jacobson}, \&
  {Placco}}]{frebel2015b}
{Frebel}, A., {Chiti}, A., {Ji}, A.~P., {Jacobson}, H.~R., \& {Placco}, V.~M.
  2015, \apjl, 810, L27

\bibitem[{{Frebel} \& {Norris}(2015)}]{frebel2015}
{Frebel}, A., \& {Norris}, J.~E. 2015, \araa, 53, 631

\bibitem[{{Frebel} {et~al.}(2014){Frebel}, {Simon}, \& {Kirby}}]{frebel2014}
{Frebel}, A., {Simon}, J.~D., \& {Kirby}, E.~N. 2014, \apj, 786, 74

\bibitem[{{Frebel} {et~al.}(2005){Frebel}, {Aoki}, {Christlieb}, {Ando},
  {Asplund}, {Barklem}, {Beers}, {Eriksson}, {Fechner}, {Fujimoto}, {Honda},
  {Kajino}, {Minezaki}, {Nomoto}, {Norris}, \& {Ryan}}]{frebel2005}
{Frebel}, A., {Aoki}, W., {Christlieb}, N., {et~al.} 2005, \nat, 434, 871

\bibitem[{{Frebel} {et~al.}(2006){Frebel}, {Christlieb}, {Norris}, {Beers},
  {Bessell}, {Rhee}, {Fechner}, {Marsteller}, {Rossi}, {Thom}, {Wisotzki}, \&
  {Reimers}}]{frebel2006}
{Frebel}, A., {Christlieb}, N., {Norris}, J.~E., {et~al.} 2006, \apj, 652, 1585

\bibitem[{{Gaia Collaboration} {et~al.}(2016){Gaia Collaboration}, {Prusti},
  {de Bruijne}, {Brown}, {Vallenari}, {Babusiaux}, {Bailer-Jones}, {Bastian},
  {Biermann}, {Evans}, \& et~al.}]{gaia2016}
{Gaia Collaboration}, {Prusti}, T., {de Bruijne}, J.~H.~J., {et~al.} 2016,
  \aap, 595, A1

\bibitem[{{Gull} {et~al.}(2018){Gull}, {Frebel}, \& {Others}}]{gull2018}
{Gull}, M., {Frebel}, A., \& {Others}. 2018, \apj, arXiv:1800.00000

\bibitem[{{Hansen} {et~al.}(2014){Hansen}, {Hansen}, {Christlieb}, {Yong},
  {Bessell}, {Garc{\'{\i}}a P{\'e}rez}, {Beers}, {Placco}, {Frebel}, {Norris},
  \& {Asplund}}]{hansen2014}
{Hansen}, T., {Hansen}, C.~J., {Christlieb}, N., {et~al.} 2014, \apj, 787, 162

\bibitem[{{Hansen} {et~al.}(2015){Hansen}, {Hansen}, {Christlieb}, {Beers},
  {Yong}, {Bessell}, {Frebel}, {Garc{\'{\i}}a P{\'e}rez}, {Placco}, {Norris},
  \& {Asplund}}]{hansen2015}
---. 2015, \apj, 807, 173

\bibitem[{{Hansen} {et~al.}(2018){Hansen}, {Holmbeck}, \& {Beers}}]{hansen2018}
{Hansen}, T.~T., {Holmbeck}, E.~M., \& {Beers}, T.~C. 2018, \apj,
  arXiv:1800.00000

\bibitem[{{Hartwig} {et~al.}(2018){Hartwig}, {Yoshida}, {Magg}, {Frebel},
  {Glover}, {G{\'o}mez}, {Griffen}, {Ishigaki}, {Ji}, {Klessen}, {O'Shea}, \&
  {Tominaga}}]{hartwig2018}
{Hartwig}, T., {Yoshida}, N., {Magg}, M., {et~al.} 2018, ArXiv e-prints,
  arXiv:1801.05044

\bibitem[{{Ito} {et~al.}(2013){Ito}, {Aoki}, {Beers}, {Tominaga}, {Honda}, \&
  {Carollo}}]{ito2013}
{Ito}, H., {Aoki}, W., {Beers}, T.~C., {et~al.} 2013, \apj, 773, 33

\bibitem[{{Jeon} {et~al.}(2017){Jeon}, {Besla}, \& {Bromm}}]{jeon2017}
{Jeon}, M., {Besla}, G., \& {Bromm}, V. 2017, \apj, 848, 85

\bibitem[{{Ji} {et~al.}(2016){Ji}, {Frebel}, {Chiti}, \& {Simon}}]{ji2016}
{Ji}, A.~P., {Frebel}, A., {Chiti}, A., \& {Simon}, J.~D. 2016, \nat, 531, 610

\bibitem[{{Keller} {et~al.}(2007){Keller}, {Schmidt}, {Bessell}, {Conroy},
  {Francis}, {Granlund}, {Kowald}, {Oates}, {Martin-Jones}, {Preston},
  {Tisserand}, \& {Vaccarella}}]{keller2007}
{Keller}, S.~C., {Schmidt}, B.~P., {Bessell}, M.~S., {et~al.} 2007, \pasa, 24,
  1

\bibitem[{{Keller} {et~al.}(2014){Keller}, {Bessell}, {Frebel}, {Casey},
  {Asplund}, {Jacobson}, {Lind}, {Norris}, {Yong}, {Heger}, {Magic}, {da
  Costa}, {Schmidt}, \& {Tisserand}}]{keller2014}
{Keller}, S.~C., {Bessell}, M.~S., {Frebel}, A., {et~al.} 2014, \nat, 506, 463

\bibitem[{{Kordopatis} {et~al.}(2013){Kordopatis}, {Gilmore}, {Steinmetz},
  {Boeche}, {Seabroke}, {Siebert}, {Zwitter}, {Binney}, {de Laverny},
  {Recio-Blanco}, {Williams}, {Piffl}, {Enke}, {Roeser}, {Bijaoui}, {Wyse},
  {Freeman}, {Munari}, {Carrillo}, {Anguiano}, {Burton}, {Campbell}, {Cass},
  {Fiegert}, {Hartley}, {Parker}, {Reid}, {Ritter}, {Russell}, {Stupar},
  {Watson}, {Bienaym{\'e}}, {Bland-Hawthorn}, {Gerhard}, {Gibson}, {Grebel},
  {Helmi}, {Navarro}, {Conrad}, {Famaey}, {Faure}, {Just}, {Kos}, {Matijevi{\v
  c}}, {McMillan}, {Minchev}, {Scholz}, {Sharma}, {Siviero}, {de Boer}, \& {{\v
  Z}erjal}}]{kordopatis2013}
{Kordopatis}, G., {Gilmore}, G., {Steinmetz}, M., {et~al.} 2013, \aj, 146, 134

\bibitem[{{Kunder} {et~al.}(2017){Kunder}, {Kordopatis}, {Steinmetz},
  {Zwitter}, {McMillan}, {Casagrande}, {Enke}, {Wojno}, {Valentini},
  {Chiappini}, {Matijevi{\v c}}, {Siviero}, {de Laverny}, {Recio-Blanco},
  {Bijaoui}, {Wyse}, {Binney}, {Grebel}, {Helmi}, {Jofre}, {Antoja}, {Gilmore},
  {Siebert}, {Famaey}, {Bienaym{\'e}}, {Gibson}, {Freeman}, {Navarro},
  {Munari}, {Seabroke}, {Anguiano}, {{\v Z}erjal}, {Minchev}, {Reid},
  {Bland-Hawthorn}, {Kos}, {Sharma}, {Watson}, {Parker}, {Scholz}, {Burton},
  {Cass}, {Hartley}, {Fiegert}, {Stupar}, {Ritter}, {Hawkins}, {Gerhard},
  {Chaplin}, {Davies}, {Elsworth}, {Lund}, {Miglio}, \& {Mosser}}]{kunder2017}
{Kunder}, A., {Kordopatis}, G., {Steinmetz}, M., {et~al.} 2017, \aj, 153, 75

\bibitem[{{Lattimer} \& {Schramm}(1974)}]{lattimer1974}
{Lattimer}, J.~M., \& {Schramm}, D.~N. 1974, \apjl, 192, L145

\bibitem[{{Lee} {et~al.}(2008{\natexlab{a}}){Lee}, {Beers}, {Sivarani},
  {Allende Prieto}, {Koesterke}, {Wilhelm}, {Re Fiorentin}, {Bailer-Jones},
  {Norris}, {Rockosi}, {Yanny}, {Newberg}, {Covey}, {Zhang}, \&
  {Luo}}]{lee2008a}
{Lee}, Y.~S., {Beers}, T.~C., {Sivarani}, T., {et~al.} 2008{\natexlab{a}}, \aj,
  136, 2022

\bibitem[{{Lee} {et~al.}(2008{\natexlab{b}}){Lee}, {Beers}, {Sivarani},
  {Johnson}, {An}, {Wilhelm}, {Allende Prieto}, {Koesterke}, {Re Fiorentin},
  {Bailer-Jones}, {Norris}, {Yanny}, {Rockosi}, {Newberg}, {Cudworth}, \&
  {Pan}}]{lee2008b}
---. 2008{\natexlab{b}}, \aj, 136, 2050

\bibitem[{{Lee} {et~al.}(2013){Lee}, {Beers}, {Masseron}, {Plez}, {Rockosi},
  {Sobeck}, {Yanny}, {Lucatello}, {Sivarani}, {Placco}, \& {Carollo}}]{lee2013}
{Lee}, Y.~S., {Beers}, T.~C., {Masseron}, T., {et~al.} 2013, \aj, 146, 132

\bibitem[{{Lindegren} {et~al.}(2016){Lindegren}, {Lammers}, {Bastian},
  {Hern{\'a}ndez}, {Klioner}, {Hobbs}, {Bombrun}, {Michalik}, {Ramos-Lerate},
  {Butkevich}, {Comoretto}, {Joliet}, {Holl}, {Hutton}, {Parsons},
  {Steidelm{\"u}ller}, {Abbas}, {Altmann}, {Andrei}, {Anton}, {Bach},
  {Barache}, {Becciani}, {Berthier}, {Bianchi}, {Biermann}, {Bouquillon},
  {Bourda}, {Br{\"u}semeister}, {Bucciarelli}, {Busonero}, {Carlucci},
  {Casta{\~n}eda}, {Charlot}, {Clotet}, {Crosta}, {Davidson}, {de Felice},
  {Drimmel}, {Fabricius}, {Fienga}, {Figueras}, {Fraile}, {Gai}, {Garralda},
  {Geyer}, {Gonz{\'a}lez-Vidal}, {Guerra}, {Hambly}, {Hauser}, {Jordan},
  {Lattanzi}, {Lenhardt}, {Liao}, {L{\"o}ffler}, {McMillan}, {Mignard}, {Mora},
  {Morbidelli}, {Portell}, {Riva}, {Sarasso}, {Serraller}, {Siddiqui}, {Smart},
  {Spagna}, {Stampa}, {Steele}, {Taris}, {Torra}, {van Reeven}, {Vecchiato},
  {Zschocke}, {de Bruijne}, {Gracia}, {Raison}, {Lister}, {Marchant},
  {Messineo}, {Soffel}, {Osorio}, {de Torres}, \& {O'Mullane}}]{lindegren2016}
{Lindegren}, L., {Lammers}, U., {Bastian}, U., {et~al.} 2016, \aap, 595, A4

\bibitem[{{Matijevi{\v c}} {et~al.}(2012){Matijevi{\v c}}, {Zwitter},
  {Bienaym{\'e}}, {Bland-Hawthorn}, {Boeche}, {Freeman}, {Gibson}, {Gilmore},
  {Grebel}, {Helmi}, {Munari}, {Navarro}, {Parker}, {Reid}, {Seabroke},
  {Siebert}, {Siviero}, {Steinmetz}, {Watson}, {Williams}, \&
  {Wyse}}]{matijevic2012}
{Matijevi{\v c}}, G., {Zwitter}, T., {Bienaym{\'e}}, O., {et~al.} 2012, \apjs,
  200, 14

\bibitem[{Mcmahon(1979)}]{sed}
Mcmahon, L.~E. 1979, in UNIX Programmer’s Manual - 7th Edition, volume 2,
  Bell Telephone Laboratories (Murray Hill)

\bibitem[{{Ness} {et~al.}(2015){Ness}, {Hogg}, {Rix}, {Ho}, \&
  {Zasowski}}]{ness2015}
{Ness}, M., {Hogg}, D.~W., {Rix}, H.-W., {Ho}, A.~Y.~Q., \& {Zasowski}, G.
  2015, \apj, 808, 16

\bibitem[{{Norris} {et~al.}(2013){Norris}, {Bessell}, {Yong}, {Christlieb},
  {Barklem}, {Asplund}, {Murphy}, {Beers}, {Frebel}, \& {Ryan}}]{norris2013}
{Norris}, J.~E., {Bessell}, M.~S., {Yong}, D., {et~al.} 2013, \apj, 762, 25

\bibitem[{{Placco} {et~al.}(2016{\natexlab{a}}){Placco}, {Beers}, {Reggiani},
  \& {Mel{\'e}ndez}}]{placco2016}
{Placco}, V.~M., {Beers}, T.~C., {Reggiani}, H., \& {Mel{\'e}ndez}, J.
  2016{\natexlab{a}}, \apjl, 829, L24

\bibitem[{{Placco} {et~al.}(2014{\natexlab{a}}){Placco}, {Frebel}, {Beers},
  {Christlieb}, {Lee}, {Kennedy}, {Rossi}, \& {Santucci}}]{placco2014}
{Placco}, V.~M., {Frebel}, A., {Beers}, T.~C., {et~al.} 2014{\natexlab{a}},
  \apj, 781, 40

\bibitem[{{Placco} {et~al.}(2013){Placco}, {Frebel}, {Beers}, {Karakas},
  {Kennedy}, {Rossi}, {Christlieb}, \& {Stancliffe}}]{placco2013}
---. 2013, \apj, 770, 104

\bibitem[{{Placco} {et~al.}(2014{\natexlab{b}}){Placco}, {Frebel}, {Beers}, \&
  {Stancliffe}}]{placco2014c}
{Placco}, V.~M., {Frebel}, A., {Beers}, T.~C., \& {Stancliffe}, R.~J.
  2014{\natexlab{b}}, \apj, 797, 21

\bibitem[{{Placco} {et~al.}(2015{\natexlab{a}}){Placco}, {Frebel}, {Lee},
  {Jacobson}, {Beers}, {Pena}, {Chan}, \& {Heger}}]{placco2015}
{Placco}, V.~M., {Frebel}, A., {Lee}, Y.~S., {et~al.} 2015{\natexlab{a}}, \apj,
  809, 136

\bibitem[{{Placco} {et~al.}(2010){Placco}, {Kennedy}, {Rossi}, {Beers}, {Lee},
  {Christlieb}, {Sivarani}, {Reimers}, \& {Wisotzki}}]{placco2010}
{Placco}, V.~M., {Kennedy}, C.~R., {Rossi}, S., {et~al.} 2010, \aj, 139, 1051

\bibitem[{{Placco} {et~al.}(2011){Placco}, {Kennedy}, {Beers}, {Christlieb},
  {Rossi}, {Sivarani}, {Lee}, {Reimers}, \& {Wisotzki}}]{placco2011}
{Placco}, V.~M., {Kennedy}, C.~R., {Beers}, T.~C., {et~al.} 2011, \aj, 142, 188

\bibitem[{{Placco} {et~al.}(2014{\natexlab{c}}){Placco}, {Beers}, {Roederer},
  {Cowan}, {Frebel}, {Filler}, {Ivans}, {Lawler}, {Schatz}, {Sneden}, {Sobeck},
  {Aoki}, \& {Smith}}]{placco2014b}
{Placco}, V.~M., {Beers}, T.~C., {Roederer}, I.~U., {et~al.}
  2014{\natexlab{c}}, \apj, 790, 34

\bibitem[{{Placco} {et~al.}(2015{\natexlab{b}}){Placco}, {Beers}, {Ivans},
  {Filler}, {Imig}, {Roederer}, {Abate}, {Hansen}, {Cowan}, {Frebel}, {Lawler},
  {Schatz}, {Sneden}, {Sobeck}, {Aoki}, {Smith}, \& {Bolte}}]{placco2015b}
{Placco}, V.~M., {Beers}, T.~C., {Ivans}, I.~I., {et~al.} 2015{\natexlab{b}},
  \apj, 812, 109

\bibitem[{{Placco} {et~al.}(2016{\natexlab{b}}){Placco}, {Frebel}, {Beers},
  {Yoon}, {Chiti}, {Heger}, {Chan}, {Casey}, \& {Christlieb}}]{placco2016b}
{Placco}, V.~M., {Frebel}, A., {Beers}, T.~C., {et~al.} 2016{\natexlab{b}},
  \apj, 833, 21

\bibitem[{{Placco} {et~al.}(2017){Placco}, {Holmbeck}, {Frebel}, {Beers},
  {Surman}, {Ji}, {Ezzeddine}, {Points}, {Kaleida}, {Hansen}, {Sakari}, \&
  {Casey}}]{placco2017}
{Placco}, V.~M., {Holmbeck}, E.~M., {Frebel}, A., {et~al.} 2017, \apj, 844, 18

\bibitem[{{R Core Team}(2015)}]{rproject}
{R Core Team}. 2015, R: A Language and Environment for Statistical Computing, R
  Foundation for Statistical Computing, Vienna, Austria

\bibitem[{{Recio-Blanco} {et~al.}(2014){Recio-Blanco}, {de Laverny},
  {Kordopatis}, {Helmi}, {Hill}, {Gilmore}, {Wyse}, {Adibekyan}, {Randich},
  {Asplund}, {Feltzing}, {Jeffries}, {Micela}, {Vallenari}, {Alfaro}, {Allende
  Prieto}, {Bensby}, {Bragaglia}, {Flaccomio}, {Koposov}, {Korn}, {Lanzafame},
  {Pancino}, {Smiljanic}, {Jackson}, {Lewis}, {Magrini}, {Morbidelli},
  {Prisinzano}, {Sacco}, {Worley}, {Hourihane}, {Bergemann}, {Costado},
  {Heiter}, {Joffre}, {Lardo}, {Lind}, \& {Maiorca}}]{recio-blanco2014}
{Recio-Blanco}, A., {de Laverny}, P., {Kordopatis}, G., {et~al.} 2014, \aap,
  567, A5

\bibitem[{{Roederer}(2012)}]{roederer2012c}
{Roederer}, I.~U. 2012, \apj, 756, 36

\bibitem[{{Roederer} {et~al.}(2014{\natexlab{a}}){Roederer}, {Cowan},
  {Preston}, {Shectman}, {Sneden}, \& {Thompson}}]{roederer2014d}
{Roederer}, I.~U., {Cowan}, J.~J., {Preston}, G.~W., {et~al.}
  2014{\natexlab{a}}, \mnras, 445, 2970

\bibitem[{{Roederer} {et~al.}(2016{\natexlab{a}}){Roederer}, {Placco}, \&
  {Beers}}]{roederer2016}
{Roederer}, I.~U., {Placco}, V.~M., \& {Beers}, T.~C. 2016{\natexlab{a}},
  \apjl, 824, L19

\bibitem[{{Roederer} {et~al.}(2014{\natexlab{b}}){Roederer}, {Preston},
  {Thompson}, {Shectman}, {Sneden}, {Burley}, \& {Kelson}}]{roederer2014}
{Roederer}, I.~U., {Preston}, G.~W., {Thompson}, I.~B., {et~al.}
  2014{\natexlab{b}}, \aj, 147, 136

\bibitem[{{Roederer} {et~al.}(2012){Roederer}, {Lawler}, {Sobeck}, {Beers},
  {Cowan}, {Frebel}, {Ivans}, {Schatz}, {Sneden}, \&
  {Thompson}}]{roederer2012d}
{Roederer}, I.~U., {Lawler}, J.~E., {Sobeck}, J.~S., {et~al.} 2012, \apjs, 203,
  27

\bibitem[{{Roederer} {et~al.}(2014{\natexlab{c}}){Roederer}, {Schatz},
  {Lawler}, {Beers}, {Cowan}, {Frebel}, {Ivans}, {Sneden}, \&
  {Sobeck}}]{roederer2014c}
{Roederer}, I.~U., {Schatz}, H., {Lawler}, J.~E., {et~al.} 2014{\natexlab{c}},
  \apj, 791, 32

\bibitem[{{Roederer} {et~al.}(2016{\natexlab{b}}){Roederer}, {Mateo}, {Bailey},
  {Song}, {Bell}, {Crane}, {Loebman}, {Nidever}, {Olszewski}, {Shectman},
  {Thompson}, {Valluri}, \& {Walker}}]{roederer2016b}
{Roederer}, I.~U., {Mateo}, M., {Bailey}, III, J.~I., {et~al.}
  2016{\natexlab{b}}, \aj, 151, 82

\bibitem[{{Rossi} {et~al.}(1999){Rossi}, {Beers}, \& {Sneden}}]{rossi1999}
{Rossi}, S., {Beers}, T.~C., \& {Sneden}, C. 1999, in Astronomical Society of
  the Pacific Conference Series, Vol. 165, The Third Stromlo Symposium: The
  Galactic Halo, ed. B.~K. {Gibson}, R.~S. {Axelrod}, \& M.~E. {Putman}, 264

\bibitem[{{Rossi} {et~al.}(2005){Rossi}, {Beers}, {Sneden}, {Sevastyanenko},
  {Rhee}, \& {Marsteller}}]{rossi2005}
{Rossi}, S., {Beers}, T.~C., {Sneden}, C., {et~al.} 2005, \aj, 130, 2804

\bibitem[{{Sakari} {et~al.}(2018){Sakari}, {Placco}, {Hansen}, {Holmbeck},
  {Beers}, {Frebel}, {Roederer}, {Venn}, {Wallerstein}, {Davis}, {Farrell}, \&
  {Yong}}]{sakari2018}
{Sakari}, C.~M., {Placco}, V.~M., {Hansen}, T., {et~al.} 2018, \apjl, 854, L20

\bibitem[{{Schlaufman} \& {Casey}(2014)}]{schlaufman2014}
{Schlaufman}, K.~C., \& {Casey}, A.~R. 2014, \apj, 797, 13

\bibitem[{{Schlegel} {et~al.}(1998){Schlegel}, {Finkbeiner}, \&
  {Davis}}]{schlegel1998}
{Schlegel}, D.~J., {Finkbeiner}, D.~P., \& {Davis}, M. 1998, \apj, 500, 525

\bibitem[{{Shappee} {et~al.}(2017){Shappee}, {Simon}, {Drout}, {Piro},
  {Morrell}, {Prieto}, {Kasen}, {Holoien}, {Kollmeier}, {Kelson}, {Coulter},
  {Foley}, {Kilpatrick}, {Siebert}, {Madore}, {Murguia-Berthier}, {Pan},
  {Prochaska}, {Ramirez-Ruiz}, {Rest}, {Adams}, {Alatalo}, {Ba{\~n}ados},
  {Baughman}, {Bernstein}, {Bitsakis}, {Boutsia}, {Bravo}, {Di Mille}, {Higgs},
  {Ji}, {Maravelias}, {Marshall}, {Placco}, {Prieto}, \& {Wan}}]{shappee2017}
{Shappee}, B.~J., {Simon}, J.~D., {Drout}, M.~R., {et~al.} 2017, Science, 358,
  1574

\bibitem[{{Sharma} {et~al.}(2017){Sharma}, {Theuns}, \& {Frenk}}]{sharma2017}
{Sharma}, M., {Theuns}, T., \& {Frenk}, C. 2017, ArXiv e-prints,
  arXiv:1712.05811

\bibitem[{{Skrutskie} {et~al.}(2006){Skrutskie}, {Cutri}, {Stiening},
  {Weinberg}, {Schneider}, {Carpenter}, {Beichman}, {Capps}, {Chester}, \&
  {Elias}}]{skrutskie2006}
{Skrutskie}, M.~F., {Cutri}, R.~M., {Stiening}, R., {et~al.} 2006, \aj, 131,
  1163

\bibitem[{{Sneden} {et~al.}(1998){Sneden}, {Cowan}, {Burris}, \&
  {Truran}}]{sneden1998}
{Sneden}, C., {Cowan}, J.~J., {Burris}, D.~L., \& {Truran}, J.~W. 1998, \apj,
  496, 235

\bibitem[{{Sneden} {et~al.}(1994){Sneden}, {Preston}, {McWilliam}, \&
  {Searle}}]{sneden1994}
{Sneden}, C., {Preston}, G.~W., {McWilliam}, A., \& {Searle}, L. 1994, \apjl,
  431, L27

\bibitem[{{Steinmetz} {et~al.}(2006){Steinmetz}, {Zwitter}, {Siebert},
  {Watson}, {Freeman}, {Munari}, {Campbell}, {Williams}, {Seabroke}, {Wyse},
  {Parker}, {Bienaym{\'e}}, {Roeser}, {Gibson}, {Gilmore}, {Grebel}, {Helmi},
  {Navarro}, {Burton}, {Cass}, {Dawe}, {Fiegert}, {Hartley}, {Russell},
  {Saunders}, {Enke}, {Bailin}, {Binney}, {Bland-Hawthorn}, {Boeche}, {Dehnen},
  {Eisenstein}, {Evans}, {Fiorucci}, {Fulbright}, {Gerhard}, {Jauregi}, {Kelz},
  {Mijovi{\'c}}, {Minchev}, {Parmentier}, {Pe{\~n}arrubia}, {Quillen}, {Read},
  {Ruchti}, {Scholz}, {Siviero}, {Smith}, {Sordo}, {Veltz}, {Vidrih}, {von
  Berlepsch}, {Boyle}, \& {Schilbach}}]{steinmetz2006}
{Steinmetz}, M., {Zwitter}, T., {Siebert}, A., {et~al.} 2006, \aj, 132, 1645

\bibitem[{{Tody}(1986)}]{tody1986}
{Tody}, D. 1986, in \procspie, Vol. 627, Instrumentation in astronomy VI, ed.
  D.~L. {Crawford}, 733

\bibitem[{{Tody}(1993)}]{tody1993}
{Tody}, D. 1993, in Astronomical Society of the Pacific Conference Series,
  Vol.~52, Astronomical Data Analysis Software and Systems II, ed. R.~J.
  {Hanisch}, R.~J.~V. {Brissenden}, \& J.~{Barnes}, 173

\bibitem[{Williams \& Kelley(2015)}]{gnuplot}
Williams, T., \& Kelley, C. 2015, Gnuplot 5.0: an interactive plotting program,
  \href{http://www.gnuplot.info/}{http://www.gnuplot.info/}

\bibitem[{Wilson(1927)}]{wilson1927}
Wilson, E.~B. 1927, Journal of the American Statistical Association, 22, 209

\bibitem[{{Yanny} {et~al.}(2009){Yanny}, {Rockosi}, {Newberg}, {Knapp},
  {Adelman-McCarthy}, {Alcorn}, {Allam}, {Allende Prieto}, {An}, {Anderson},
  {Anderson}, \& {Bailer-Jones}}]{yanny2009}
{Yanny}, B., {Rockosi}, C., {Newberg}, H.~J., {et~al.} 2009, \aj, 137, 4377

\bibitem[{{Yong} {et~al.}(2013){Yong}, {Norris}, {Bessell}, {Christlieb},
  {Asplund}, {Beers}, {Barklem}, {Frebel}, \& {Ryan}}]{yong2013}
{Yong}, D., {Norris}, J.~E., {Bessell}, M.~S., {et~al.} 2013, \apj, 762, 26

\bibitem[{{Yoon} {et~al.}(2018){Yoon}, {Beers}, {Placco}, \&
  {Rasmussen}}]{yoon2018}
{Yoon}, J., {Beers}, T.~C., {Placco}, V.~M., \& {Rasmussen}, K.~C. 2018, \apj,
  833, 20

\bibitem[{{Yoon} {et~al.}(2016){Yoon}, {Beers}, {Placco}, {Rasmussen},
  {Carollo}, {He}, {Hansen}, {Roederer}, \& {Zeanah}}]{yoon2016}
{Yoon}, J., {Beers}, T.~C., {Placco}, V.~M., {et~al.} 2016, \apj, 833, 20

\bibitem[{{York} {et~al.}(2000){York}, {Adelman}, {Anderson}, {Anderson},
  {Annis}, {Bahcall}, {Bakken}, {Barkhouser}, {Bastian}, {Berman}, {Boroski},
  {Bracker}, \& {Briegel}}]{york2000}
{York}, D.~G., {Adelman}, J., {Anderson}, Jr., J.~E., {et~al.} 2000, \aj, 120,
  1579

\bibitem[{{Zwitter} {et~al.}(2008){Zwitter}, {Siebert}, {Munari}, {Freeman},
  {Siviero}, {Watson}, {Fulbright}, {Wyse}, {Campbell}, {Seabroke}, {Williams},
  {Steinmetz}, {Bienaym{\'e}}, {Gilmore}, {Grebel}, {Helmi}, {Navarro},
  {Anguiano}, {Boeche}, {Burton}, {Cass}, {Dawe}, {Fiegert}, {Hartley},
  {Russell}, {Veltz}, {Bailin}, {Binney}, {Bland-Hawthorn}, {Brown}, {Dehnen},
  {Evans}, {Re Fiorentin}, {Fiorucci}, {Gerhard}, {Gibson}, {Kelz}, {Kujken},
  {Matijevi{\v c}}, {Minchev}, {Parker}, {Pe{\~n}arrubia}, {Quillen}, {Read},
  {Reid}, {Roeser}, {Ruchti}, {Scholz}, {Smith}, {Sordo}, {Tolstoi},
  {Tomasella}, {Vidrih}, \& {Wylie-de Boer}}]{zwitter2008}
{Zwitter}, T., {Siebert}, A., {Munari}, U., {et~al.} 2008, \aj, 136, 421

\end{thebibliography}

\clearpage
\startlongtable

\begin{deluxetable*}{@{}cclllr}
\tabletypesize{\scriptsize}
\tabletypesize{\tiny}
\tablewidth{0pc}
\tablecaption{Observing Details \label{telescope}}
\tablehead{
\colhead{Star Name}   &
\colhead{Date}        &
\colhead{Telescope}   &
\colhead{Instrument}  &
\colhead{Proposal ID} &
\colhead{Exp.}        \\
\colhead{(RAVE)}      &
\colhead{(UTC)}       &
\colhead{}            &
\colhead{}            &
\colhead{}            &
\colhead{(s)}         }
\startdata
J000022.6$-$130228 &   2014$-$09$-$11 &          Mayall &     RCSPEC &        14B$-$0231 &  720 \\
J000024.2$-$110746 &   2014$-$09$-$12 &          Mayall &     RCSPEC &        14B$-$0231 &  600 \\
J000245.9$-$325734 &   2015$-$08$-$21 &        Gemini$-$S &     GMOS$-$S &  GS$-$2015A$-$Q$-$205 &  720 \\
J000351.1$-$123154 &   2014$-$09$-$12 &          Mayall &     RCSPEC &        14B$-$0231 &  720 \\
J000414.3$-$265709 &   2014$-$10$-$29 &            SOAR &    Goodman &      2014B$-$0231 &   30 \\
J000415.8$-$581553 &   2015$-$08$-$31 &            SOAR &    Goodman &      2015B$-$0020 &   60 \\
J000417.4$-$124503 &   2014$-$09$-$11 &          Mayall &     RCSPEC &        14B$-$0231 &  720 \\
J000438.4$-$540134 &   2014$-$12$-$26 &            SOAR &    Goodman &      2014B$-$0231 &  120 \\
\enddata
\end{deluxetable*}

\clearpage
\startlongtable

\begin{deluxetable*}{cccrrrrrrr}
\tabletypesize{\scriptsize}
\tabletypesize{\tiny}
\tablewidth{0pc}
\tablecaption{Coordinates, Magnitudes, Color Indices, and Redenning Estimates \label{coordsmags}}
\tablehead{
\colhead{Star Name} &
\colhead{$\alpha$}  &
\colhead{$\delta$}  &
\colhead{$l$}       &
\colhead{$b$}       &
\colhead{$V$}       &
\colhead{$(B-V)$}   &
\colhead{$J$}       &
\colhead{$(J-K)$}   &
\colhead{$E(B-V)$}  \\
\colhead{(RAVE)}    &
\colhead{(J2000)}   &
\colhead{(J2000)}   &
\colhead{(deg)}     &
\colhead{(deg)}     &
\colhead{(mag)}     &
\colhead{}          &
\colhead{(mag)}     &
\colhead{}          &
\colhead{}          }
\startdata
\input{table2c.tex}
\enddata
\end{deluxetable*}

\clearpage
\startlongtable

\begin{longrotatetable}
\begin{deluxetable*}{@{}c@{}rcrrc@{}rccrrc@{}rcrrc@{}crrrrr}
\movetabledown=0.40in
\tabletypesize{\scriptsize}
\tabletypesize{\tiny}
\tablewidth{0pc}
\tablecaption{Stellar Parameters from RAVE DR4, RAVE DR5, RAVE-on, and the n-SSPP \label{comprave}}
\tablehead{
\colhead{}                    &
\multicolumn{4}{c}{RAVE DR4}  &
\colhead{}                    &
\multicolumn{5}{c}{RAVE DR5}  &
\colhead{}                    &
\multicolumn{4}{c}{RAVE-on}   &
\colhead{}                    &
\multicolumn{6}{c}{This work} \\
\cline{2-5} \cline{7-11} 
\cline{13-16} \cline{18-23}   \\
\colhead{Star Name}           &
\colhead{$V_r$}               &
\colhead{\teff}               &
\colhead{\logg}               &
\colhead{\metal}              &&
\colhead{$V_r$}               &
\colhead{\teff}               &
\colhead{\teff$_{\rm IR}$}    &
\colhead{\logg}               &
\colhead{\metal}              &&
\colhead{$V_r$}               &
\colhead{\teff}               &
\colhead{\logg}               &
\colhead{\metal}              &&
\colhead{\teff}               &
\colhead{\logg}               &
\colhead{\metal}              &
\colhead{\cfe}                &
\colhead{$\Delta$\cfe}        &
\colhead{\afe}                \\
\colhead{(RAVE)}              &
\colhead{(km~s$^{-1}$)}       &
\colhead{(K)}                 &
\colhead{(cgs)}               &
\colhead{}                    &&
\colhead{(km~s$^{-1}$)}       &
\colhead{(K)}                 &
\colhead{(K)}                 &
\colhead{(cgs)}               &
\colhead{}                    &&
\colhead{(km~s$^{-1}$)}       &
\colhead{(K)}                 &
\colhead{(cgs)}               &
\colhead{}                    &&
\colhead{(K)}                 &
\colhead{(cgs)}               &
\colhead{}                    &
\colhead{}                    &
\colhead{}                    &
\colhead{}                    }
\startdata
\input{table3bc.tex}
\enddata
\end{deluxetable*}
\end{longrotatetable}

\clearpage
\startlongtable

\begin{deluxetable*}{@{}ccrrrrrrrrrrr}
\tabletypesize{\scriptsize}
\tabletypesize{\tiny}
\tablewidth{0pc}
\tablecaption{Gaia/TGAS parameters \label{gaiar}}
\tablehead{
\colhead{Star Name}                &
\colhead{ID   }                    &
\colhead{$G$}                      &
\colhead{$\mu_{\rm RA}$}           &
\colhead{$\sigma_{\mu_{\rm RA}}$}  &
\colhead{$\mu_{\rm DEC}$}          &
\colhead{$\sigma_{\mu_{\rm DEC}}$} &
\colhead{$\pi$}                    &
\colhead{$\sigma_{\pi}$}           &
\colhead{$D$}                      &
\colhead{$\sigma_{D}$}             &
\colhead{$V_{T}$}                  &
\colhead{$Z$}                      \\
\colhead{(RAVE)}                   &
\colhead{(GAIA)}                   &
\colhead{(mag)}                    &
\colhead{(mas yr$^{-1}$)}          &
\colhead{(mas yr$^{-1}$)}          &
\colhead{(mas yr$^{-1}$)}          &
\colhead{(mas yr$^{-1}$)}          &
\colhead{(mas)}                    &
\colhead{(mas)}                    &
\colhead{(kpc)}                    &
\colhead{(kpc)}                    &
\colhead{(km s$^{-1}$)}            &
\colhead{(kpc)}                    }
\startdata
J000245.9$-$325734 & 2313475887653200384 & 11.75 &     78.862 &      0.804 &  $-$53.267 &      0.857 &      0.600 &      0.340 &      1.108 &      0.811 &  279 & $-$0.595\\
J000414.3$-$265709 & 2333779366292490240 & 10.39 &      8.502 &      0.152 &  $-$14.463 &      0.082 &      0.380 &      0.280 &      1.454 &      1.281 &  100 & 1.185\\
J000415.8$-$581553 & 4919110843943100416 & 10.22 &     11.667 &      0.524 &   $-$7.546 &      0.535 &      0.170 &      0.250 &      2.133 &      2.027 &   78 & $-$2.015\\
J000417.4$-$124503 & 2421324921637437952 & 12.08 &     20.737 &      1.761 &  $-$48.790 &      1.004 &      0.740 &      0.540 &      0.842 &      0.715 &  210 & $-$0.314\\
J000438.4$-$540134 & 4923965909334259712 & 12.01 &     22.633 &      0.381 &  $-$13.198 &      0.397 &      0.170 &      0.220 &      2.379 &      2.264 &  255 & 2.092\\
J000721.6$-$134824 & 2417779821271766016 & 10.21 &     69.591 &      1.000 &  $-$71.637 &      0.642 &      2.170 &      0.370 &      0.454 &      0.092 &  161 & 0.359\\
J000953.8$-$612618 & 4904914499561063424 & 11.04 &      9.215 &      0.921 &   $-$7.322 &      0.606 &   $-$0.130 &      0.500 &      1.419 &      1.847 &   50 & 1.419\\
J001236.5$-$181631 & 2366298075518998016 & 10.69 &  $-$30.166 &      1.036 &  $-$61.608 &      0.684 &      1.550 &      0.300 &      0.620 &      0.142 &  195 & $-$0.558\\
\enddata
\end{deluxetable*}

\clearpage
\startlongtable

\begin{deluxetable*}{cccccccr}
\tabletypesize{\scriptsize}
\tabletypesize{\tiny}
\tablewidth{0pc}
\tablecaption{Group Assignments for CEMP Stars\label{groups}}
\tablehead{
\colhead{Star Name}      &
\colhead{$V$}            &
\colhead{\teff}          &
\colhead{\metal}         &
\colhead{\cfe}           &
\colhead{\cfe$_{\rm c}$} &
\colhead{A(C)$_{\rm c}$} &
\colhead{Group}          \\
\colhead{(RAVE)}         &
\colhead{(mag)}          &
\colhead{(K)}            &
\colhead{}               &
\colhead{}               &
\colhead{}               &
\colhead{}               &
\colhead{}               }
\startdata
J001548.1$-$625321 &   11.04 & 4711 & $-$2.63 & $-$0.01 & $+$0.70 &  6.50 &  II \\
J002330.7$-$163143 &   12.28 & 5542 & $-$2.40 & $+$0.76 & $+$0.77 &  6.80 &   I \\
J004539.3$-$745729 &   11.67 & 5049 & $-$2.24 & $+$1.24 & $+$1.37 &  7.56 &   I \\
J005419.7$-$061155 &   13.07 & 4897 & $-$2.59 & $+$0.30 & $+$0.72 &  6.56 &  II \\
J011553.2$-$733050 &   10.81 & 6422 & $-$1.56 & $+$0.83 & $+$0.83 &  7.70 &   I \\
J012931.1$-$160046 &   11.64 & 5096 & $-$2.71 & $+$0.62 & $+$0.94 &  6.66 &  II \\
J013346.6$-$272737 &   12.40 & 4821 & $-$2.79 & $+$2.10 & $+$2.27 &  7.91 &   I \\
J014738.1$-$113048 &   12.99 & 4874 & $-$2.90 & $+$1.31 & $+$1.60 &  7.13 & III \\
\enddata
\end{deluxetable*}

\end{document}